\documentclass{aa}

\usepackage[T1]{fontenc}
\usepackage[utf8]{inputenc}
\usepackage{newtxtext,newtxmath}
\usepackage{graphicx}
\usepackage[colorlinks=true, linkcolor=blue, citecolor=blue, urlcolor=gray]{hyperref}
\usepackage{xparse}

\NewDocumentCommand{\nativename}{O{1.6ex} O{-0.25ex} m}{%
  \texorpdfstring{%
    \raisebox{#2}{%
      \includegraphics[height=#1]{#3}%
    }%
  }{}%
}

\bibpunct{(}{)}{;}{a}{}{,}

\begin{document}

\title{The TNG50-SKIRT Atlas: Spatially resolved synthetic galaxies from the ultraviolet to the submillimetre (DR2)}

\titlerunning{TNG50-SKIRT Atlas}

\author{%
Maarten~Baes\inst{\ref{UGent}}\thanks{\email{maarten.baes@ugent.be}}
\and
Paul~Vauterin\inst{\ref{UGent}}
\and
Abdurro'uf\inst{\ref{Indiana}}
\and
Nick~Andreadis\inst{\ref{UGent}}~(\nativename[1.8ex][-0.43ex]{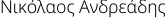})
\and
\\
Thiago~Bueno-Dalpiaz\inst{\ref{Rio}}
\and
Sena~Bokona~Tulu\inst{\ref{Jimma},\ref{UGent}}~(\nativename[1.5ex][0ex]{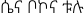})
\and
Peter~Camps\inst{\ref{UGent}}
\and
\\
Abdissa~Tassama~Emana\inst{\ref{Jimma},\ref{UGent}}~(\nativename[1.6ex][-0.15ex]{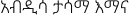})
\and
Jacopo Fritz\inst{\ref{UNAM}}
\and
Andrea~Gebek\inst{\ref{UGent}}
\and
\\
Anand~Utsav~Kapoor\inst{\ref{UGent}}~(\nativename[1.8ex][-0.43ex]{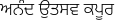})
\and
Inja~Kova\v{c}i\'{c}\inst{\ref{UGent}}
\and
Arno Lauwers\inst{\ref{UGent}}
\and
Kosei~Matsumoto\inst{\ref{UGent}}~(\nativename[1.5ex][0ex]{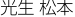})
\and
\\
Kar{\'{\i}}n~Men\'endez-Delmestre\inst{\ref{CBFP},\ref{Rio}}
\and
Aleksandr~V.~Mosenkov\inst{\ref{BYU}}~(\nativename[1.8ex][-0.42ex]{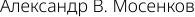})
\and
\\
Angelos~Nersesian\inst{\ref{UGent},\ref{Liege}}~(\nativename[1.8ex][-0.43ex]{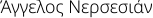})
\and 
Lara~Pantoni\inst{\ref{UGent}}
\and 
Mariana Rivas S\'anchez\inst{\ref{UNAM},\ref{UGent}}
\and
\\
Waad~Saftly\inst{\ref{Homs},\ref{UGent}}~(\nativename[1.8ex][-0.43ex]{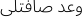})
\and
Samir~Salim\inst{\ref{Indiana}}
\and
Marko~Stalevski\inst{\ref{AOB},\ref{UGent}}
\and
Qi~Zeng\inst{\ref{SAO},\ref{UCAS},\ref{UGent}}~(\nativename[1.6ex][-0.1ex]{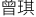})
}

\institute{%
Department of Physics and Astronomy, Universiteit Gent, Proeftuinstraat 86 N3, B-9000 Gent, Belgium
\label{UGent}
\and
Department of Astronomy, Indiana University, Bloomington, IN 47405, USA
\label{Indiana}
\and
Universidade Federal do Rio de Janeiro, Observat\'orio do Valongo, Ladeira Pedro Ant\^onio 43, Sa\'ude CEP, 20080-090 Rio de Janeiro, RJ, Brazil
\label{Rio}
\and
Physics Department, College of Natural Sciences, Jimma University, PO Box 378, Jimma, Ethiopia
\label{Jimma}
\and
Instituto de Radioastronom{\'{\i}}a y Astrof{\'{\i}}sica, Universidad Nacional Aut{\'{o}}noma de M{\'{e}}xico, Morelia, Michoac{\'{a}}n 58089, Mexico
\label{UNAM}
\and
Centro Brasiliero de Pesquisas F\'{\i}sicas (CBPF), Rua Dr. Xavier Sigaud 150, Botafogo, 22290-180 Rio de Janeiro, RJ, Brazil
\label{CBFP}
\and
Department of Physics and Astronomy, N283 ESC, Brigham Young University, Provo, UT 84602, USA
\label{BYU}
\and
STAR Institute, Universit\'e de Li\`ege, Quartier Agora, All\'ee du Six Ao\^ut 19c, B-4000 Li\`ege, Belgium
\label{Liege}
\and
Department of Mathematics, Faculty of Science, Homs University, Homs, Syria
\label{Homs}
\and
Astronomical Observatory, Volgina 7, 11060 Belgrade, Serbia
\label{AOB}
\and
Shanghai Astronomical Observatory, Chinese Academy of Sciences, No.~80 Nandan Road, Shanghai 200030, People's Republic of China
\label{SAO}
\and
School of Astronomy and Space Sciences, University of Chinese Academy of Sciences, No.~19A Yuquan Road, Beijing 100049, People's Republic of China
\label{UCAS}
}

\authorrunning{M. Baes et al.}

\date{\today}

\abstract{We present the second data release (DR2) of the TNG50-SKIRT Atlas (TSA), a library of synthetic, spatially resolved galaxy observables. The atlas is constructed by post-processing a stellar-mass-complete ($10^{9.8}~{\text{M}}_\odot < M_\star < 10^{12}~{\text{M}}_\odot$) sample of 1154 $z=0$ galaxies from the TNG50 cosmological hydrodynamical simulation with the Monte Carlo radiative transfer code {\texttt{SKIRT}}. Compared to the first release, TSA DR2 extends the wavelength coverage from the ultraviolet to the submillimetre, including dust emission, and incorporates updated stellar population models together with an improved treatment of dust-enshrouded star-forming regions. The atlas provides spatially resolved spectral energy distributions, broadband images, and physical property maps for multiple viewing orientations, as well as a catalogue of integrated properties enabling direct comparison with unresolved observations. We validate the data products through extensive quality control, including an assessment of Monte Carlo noise, and demonstrate their internal consistency using diagnostic relations between luminosities and star formation rates. TSA DR2 provides a versatile resource for studies of dust attenuation and emission, star formation tracers, galaxy morphology, and multi-wavelength scaling relations across spatial scales. The atlas and associated data products are publicly released and are intended to support a wide range of observationally oriented studies of galaxy evolution.}


\maketitle

\section{Introduction}
\label{Introduction.sec}

Cosmological hydrodynamical simulations have become indispensable tools for studying galaxy formation and evolution. Modern simulations follow the coupled evolution of dark matter, gas, stars, and black holes while incorporating key physical processes such as star formation, chemical enrichment, and feedback from supernovae and active galactic nuclei \citep{Somerville2015, Vogelsberger2020a, Crain2023}. Over the past decade, advances in numerical techniques, resolution, and subgrid modelling have enabled simulations to reproduce a wide range of observed global galaxy properties, including stellar mass functions, colour bimodality, star-formation scaling relations, and mass--metallicity relations.

To enable insightful comparison between simulations and observations, simulated galaxies must be forward-modelled into the observable domain. This requires the generation of synthetic images that account not only for stellar emission but also for the effects of absorption, scattering, and re-emission by interstellar dust. Dust attenuates a substantial fraction of stellar radiation in typical galaxies \citep{Popescu2002, Viaene2016, Bianchi2018}, and radiative transfer calculations have shown that its impact on observed fluxes, colours, and morphology is highly non-trivial and strongly geometry-dependent \citep[e.g.][]{Witt1992a, Byun1994, Pierini2004, Mollenhoff2006, Gadotti2010}. Three-dimensional (3D) dust radiative transfer modelling \citep{Steinacker2013} is therefore essential for producing realistic synthetic data products. Over the past decade, dust radiative transfer postprocessing of simulated galaxies has been an important element in the comparison of cosmological galaxy formation simulations to observational data \citep[e.g.,][]{Torrey2015, Camps2016, Camps2018a, Camps2022, Trayford2017, RodriguezGomez2019, Baes2020b, Baes2025, Kapoor2021, Faucher2023, Bottrell2024, Gebek2024, Gebek2025, Gebek2026, Lu2026a}. 

The TNG50-SKIRT Atlas \citep[TSA,][]{Baes2024a} was introduced as a high-spatial-resolution synthetic imaging database for galaxies extracted from the $z=0$ snapshot of the TNG50 simulation \citep{Pillepich2019, Nelson2019b}. It combines the statistics of large-volume simulations and the resolution of zoom-in simulations. The first data release (DR1) of the TSA presented dust-attenuated and dust-free broadband images from the ultraviolet (UV) to the near-infrared (NIR), together with intrinsic physical property maps, for a stellar-mass-selected sample of 1154 galaxies. Applications include investigations of the galaxy properties across the {\em{UVJ}} diagram \citep{Baes2024a}, the wavelength dependence of galaxy sizes \citep{Baes2024b} and disc scale lengths \citep{Emana2026}, multi-wavelength nonparametric galaxy morphology \citep{BokonaTulu2026}, and a comparison of methods to infer galaxy properties from broadband images \citep{EuclidCollaborationKovacic2025, EuclidCollaborationAbdurrouf2025, EuclidCollaborationNersesian2026}.

Despite its broad applicability, DR1 was necessarily limited in scope. Most importantly, it modelled only stellar emission and dust attenuation, and did not include thermal dust emission. As a result, the atlas could not be used to study galaxies across the full UV--to--submm wavelength range, which is useful to investigate infrared star-formation tracers \citep{Kennicutt2009, Hao2011, Belfiore2023}, dust scaling relations \citep{Cortese2012, Viaene2014, Galliano2021, Abdurrouf2022b}, or the global energy balance between absorbed and re-emitted radiation \citep{daCunha2008, Baes2010b, Mosenkov2018, Boquien2019}. Self-consistent modelling of dust emission requires significantly more computationally expensive radiative transfer calculations and was therefore deferred in the initial release.

Additional limitations emerged from the experience gained with DR1. The use of predefined broadband instruments restricted the atlas to a fixed filter set, complicating the generation of images for new or upcoming facilities such as \emph{Euclid}. Moreover, while random viewing angles are optimal for statistical studies, specific orientations such as face-on and edge-on views are particularly valuable for structural and inclination-dependent analyses \citep[e.g.][]{Patel2012, Trayford2017}. Finally, aspects of the stellar population and star-forming region modelling, although adequate on global scales, were found to produce overly clumpy and sometimes unrealistic UV morphologies on spatially resolved scales.

For these reasons, we have generated a substantially revised and expanded version of the atlas. In this paper, we present the second data release of the TNG50-SKIRT Atlas (TSA DR2). TSA DR2 contains the same set of 1154 galaxies as DR1, but extends the wavelength coverage from the UV to the submm by fully accounting for thermal dust emission, provides spectrally resolved data cubes that allow the construction of images in arbitrary filters, includes dedicated face-on and edge-on projections in addition to random orientations, and adopts an updated and more consistent treatment of stellar populations and star-forming regions. 

This paper is organised as follows. In Sect.~{\ref{Methods.sec}} we introduce the TNG50 simulation and the TSA sample extracted from it, and we describe the changes made to the {\texttt{SKIRT}} radiative transfer post-processing compared to DR1. In Sect.~{\ref{Presentation.sec}} we present the data products that we publicly release as DR2. In Sect.~{\ref{DR12.sec}} we make a comparison between DR1 and DR2, and in Sect.~{\ref{Uncertainties.sec}} we discuss the uncertainties on the images that we release. Finally, Sect.~{\ref{Summary.sec}} sums up and discusses caveats and possible future improvements of the TSA.

\section{Methods}
\label{Methods.sec}

\subsection{The TSA sample}

The TSA is based on the TNG50 simulation, the highest-resolution run of the IllustrisTNG suite of cosmological magneto-hydrodynamical simulations \citep{Pillepich2018a, Springel2018, Marinacci2018, Naiman2018, Nelson2018}. TNG50 follows the evolution of a periodic cubic volume of side length 51.7~comoving~Mpc from $z=127$ to the present day using the moving-mesh code \texttt{AREPO} \citep{Springel2010a}. It adopts a $\Lambda$CDM cosmology consistent with the \emph{Planck} 2015 results \citep{PlanckCollaboration2016} and includes a comprehensive galaxy formation model that accounts for gas cooling and heating, stochastic star formation, stellar evolution and chemical enrichment, feedback from supernovae and active galactic nuclei, magnetic fields, and the growth of supermassive black holes \citep{Weinberger2017, Pillepich2018a}. With a baryonic mass resolution of $8.5\times10^{4}~{\text{M}}_\odot$ and typical spatial resolution of order 70--140~pc in star-forming regions, TNG50 resolves the internal structure of galaxies at sub-kpc scales while retaining a statistically meaningful galaxy population \citep{Pillepich2019,Nelson2019b}.

From the TNG50 simulation we constructed the TNG50-SKIRT Atlas (TSA), a synthetic imaging data set for a complete, stellar-mass selected sample of galaxies at $z=0$ \citep{Baes2024a}. The TSA sample consists of all TNG50 galaxies with total stellar masses in the range $10^{9.8} \le M_\star/{\text{M}}_\odot \le 10^{12}$, yielding a total of 1154 galaxies. The sample includes both star-forming and quiescent systems, allowing the investigation of galaxy properties across the full diversity of the local galaxy population. For each galaxy, synthetic observations are generated using 3D radiative transfer calculations with the {\texttt{SKIRT}} code, enabling direct connections between intrinsic physical properties and observable quantities. This combination of high spatial resolution, broad coverage in galaxy properties, and physically motivated forward modelling makes the TSA a uniquely powerful resource for studies of galaxy structure, morphology, and multi-wavelength emission.

\subsection{Updates in the SKIRT setup compared to DR1}
\label{SKIRT_setup.sec}

The data products of both DR1 and DR2 are based on 3D dust radiative transfer simulations with the {\texttt{SKIRT}}\footnote{\url{https://skirt.ugent.be/}} Monte Carlo code \citep{Baes2003, Baes2011b, Camps2015a, Camps2020}. The overall philosophy and methodology we use here to generate mock observables for simulated galaxies largely follows the prescriptions provided in \citet{Baes2024a}, which itself was built on previous work \citep[e.g.][]{Camps2018a, Kapoor2021, Trcka2022}. In this subsection we describe the main changes and improvements compared to the TSA DR1.

\subsubsection{Primary radiation sources}
\label{Primary.sec}

The primary radiation sources in the TSA DR2 radiative transfer calculations are the stellar populations associated with each TNG50 galaxy. As in DR1, we separated old and young stellar particles, with 10~Myr as the boundary between them, and we assigned different template SEDs to them. The rationale behind this separation is that young stellar particles are assumed to still be partly embedded in their dusty birth clouds. The choice of the template SEDs in DR2 is different from DR1, however.

For stellar particles older than 10~Myr, TSA DR1 assigned simple stellar population (SSP) spectral energy distributions (SEDs) from the models of \citet{Bruzual2003}, assuming a \citet{Chabrier2003} initial mass function (IMF). In TSA DR2, we replace these SSPs with models from the Binary Population and Spectral Synthesis ({\texttt{BPASS}}) framework \citep{Eldridge2017, Stanway2018}. {\texttt{BPASS}} explicitly accounts for the effects of binary stellar evolution, which can significantly affect the ionising photon output and the spectral shape of stellar populations, particularly at young and intermediate ages. We use the {\texttt{BPASS}} library with the \citet{Chabrier2003} IMF to be consistent with the IMF used in the TNG50 simulation.

The treatment of young stellar populations and star-forming regions has been substantially revised. In TSA DR1, stellar particles younger than 10~Myr were treated as radiation sources and assigned \ion{H}{ii} region spectral templates based on the {\texttt{MAPPINGS}}~III library \citep{Groves2008}, with template parameters determined following the prescriptions of \citet{Trcka2022}. While this approach provides a statistically reasonable description of star forming regions on global scales, it results in very compact and spatially clumpy UV emission due to the limited number of stellar particles younger than 10 Myr.

In TSA DR2, we instead adopt \ion{H}{ii} region templates from the {\texttt{TODDLERS}} library \citep{Kapoor2023, Kapoor2024}, which are based on {\texttt{BPASS}} stellar population models and are therefore fully consistent with the SSPs used for the old stellar particles. Moreover, rather than associating these templates with young stellar particles, we use star-forming gas cells from the TNG50 simulation as the primary radiation sources for ongoing star formation. Concretely, we launch photon packages from each star-forming Voronoi gas cell from the simulated galaxy rather than from the individual stellar particles with ages below 10~Myr. {\texttt{SKIRT}} has the capability to read a Voronoi grid\footnote{More precisely, {\texttt{SKIRT}} reads the set of Voronoi generators and reconstructs the corresponding tessellation using the {\texttt{Voro++}} third-party library \citep{Rycroft2009}.} and use the resulting Voronoi cells directly as the source distribution. Here we adopt a simpler approach in which the emission associated with each star-forming gas cell is represented by a smoothed particle. The particle is placed at the corresponding cell generator, and its smoothing length is set to the cube root of the Voronoi cell volume. This choice ensures that the smoothing scale is comparable to the characteristic linear size of the original gas cell.

Since there are, on average, about 100 times more star-forming gas cells than individual young stellar particles, this implies that the emission is much better sampled in the spatial domain. On global, integrated scales, this change yields statistically equivalent star formation rates (SFRs) and luminosities. However, on spatially resolved scales it leads to a markedly different spatial distribution of the radiation sources, producing smoother UV morphologies that better reflect the extended nature of star-forming regions in galaxies (see Sect.~{\ref{DR12.sec}}).

\subsubsection{The dusty medium}
\label{DustyMedium.sec}

The treatment of the dusty interstellar medium constitutes one of the principal differences between TSA DR1 and DR2. In DR1, the {\texttt{SKIRT}} simulations were run in the {\tt{ExtinctionOnly}} mode, meaning that the radiative transfer calculations accounted only for dust absorption and scattering, and thermal dust emission was neglected. While this approach is sufficient for studies focused on the UV to NIR regime, it precludes any analysis of the mid-infrared (MIR), far-infrared (FIR), and submm emission, as well as investigations of the dust energy balance.

In TSA DR2, we therefore adopt a full treatment of thermal dust emission in addition to dust attenuation (the {\tt{DustEmission}} simulation mode in {\texttt{SKIRT}}). As in DR1, the diffuse dust component is described using the {\texttt{THEMIS}} dust model \citep{Jones2017}, which consists of a mixture of amorphous silicate grains and hydrocarbonaceous carbon grains and has been shown to reproduce a wide range of observed dust extinction and emission properties in nearby galaxies. The dust density distribution is derived following the same prescriptions as in DR1. In short, ISM gas cells are identified following the prescription by \citet{Torrey2012, Torrey2019}, and the dust density in an ISM cell is assumed to be proportional to the metal density, with a uniform dust-to-metal fraction $f_{\text{dust}} = 0.2$ \citep{Trcka2022}.

As in TSA DR1, the dust density distribution is represented on an adaptive octree grid, with up to 12 levels of grid refinement \citep{Saftly2013, Saftly2014}. An obvious alternative would be to use the native Voronoi mesh of the TNG50 simulation as the computational grid for the radiative transfer calculations, thereby avoiding any interpolation artefacts associated with the octree regridding. Indeed, this approach has successfully been adopted in several previous {\texttt{SKIRT}} studies based on moving-mesh simulations \citep[e.g.][]{RodriguezGomez2019, Schulz2020, Popping2022, GuzmanOrtega2023, GuzmanOrtega2025}. We nevertheless employ an adaptive octree grid for two reasons. First, photon propagation through octree grids is considerably faster than through Voronoi grids \citep{Camps2013}, an important consideration given the computational cost of the full dust emission calculations performed for TSA DR2. Second, although the native Voronoi mesh is the natural discretisation for the hydrodynamical simulation, there is no fundamental reason why it should also be the optimal discretisation for radiative transfer. Ideally, the computational grid should adapt to the complexity of the radiation field, which is not known a priori. The adaptive octree therefore represents a pragmatic compromise between computational efficiency and an accurate representation of the dusty medium.

For the calculation of the thermal dust emission, we discretise the grain size distributions of both silicate and hydrocarbon dust into ten size bins per component. The dust temperature distribution is then computed in every grid cell of the computational domain by balancing radiative heating and cooling for each grain size bin. This approach fully captures the effects of stochastic heating for small grains and temperature equilibrium for large grains, and allows the emergent dust emission spectrum to be computed consistently with the local radiation field. The methodology follows the standard implementation of dust thermal emission in {\texttt{SKIRT}}, as described in \citet{Camps2015b}.

\subsubsection{Synthetic instrument setup}
\label{Instruments.sec}

The instrumental setup in TSA DR2 builds upon that of the first data release while introducing several important extensions. As in DR1, each galaxy is observed from five random viewing directions (O1 to O5), with maximum angular distance between them. These orientations are fixed with respect to the TNG50 simulation volume, which implies they are different for each individual galaxy \citep[see][Sect. 2.4.3]{Baes2024a}. In addition, TSA DR2 includes two dedicated orientations (FO and EO) corresponding to face-on and edge-on views, respectively, which are particularly useful for structural studies and analyses of inclination-dependent effects. These orientations are defined by the stellar angular momentum vector, calculated within a radial range of 0.5 to 2 stellar half-mass radii.

A major upgrade in DR2 concerns the spectral sampling of the synthetic observations. DR1 relied on a predefined set of 22 UV--NIR broadband filters, consisting of 18 bands in the original release and four additional {\em{Euclid}} bands published subsequently in \citet{EuclidCollaborationKovacic2025}. TSA DR2 provides full spectrally resolved data cubes, spanning the entire UV to mm wavelength range in 344 wavelength bins (from 0.09~$\mu$m to 2~mm). The sampling is denser in the UV, optical, and NIR bands where stellar emission dominates. This approach enables the generation of broadband images in arbitrary filter sets through post-processing, for example using the convolution algorithms implemented in the Python Toolkit for {\texttt{SKIRT}} \citep[PTS\footnote{\url{https://skirt.ugent.be/root/_home.html}}:][]{Verstocken2020, Camps2020}. 

We kept the same pixel scale (100 pc) as in the DR1 version. However, to reduce both the computational cost of the radiative transfer simulations and the storage requirements of the atlas, the field of view of the instruments was reduced from $160\times160~\text{kpc}^2$ in DR1 to $80\times80~\text{kpc}^2$ in DR2. This choice still comfortably encloses the UV-to-submm-emitting regions of the vast majority of the galaxies in our sample: 99\% of the galaxies in the sample have a stellar half-mass radius smaller than 15~kpc. The advantage is a significant reduction of the memory footprint of the IFU data cubes. With 344 wavelength bins and $800 \times 800$ pixels, each synthetic IFU instrument still requires 840 MB. With seven IFU instruments per simulation, this adds up to 6.2 GB per simulated galaxy. 

\subsubsection{Dust-free simulations and probes}
\label{Probes.sec}

One of the main goals of the TSA is to investigate the effect of dust attenuation on the observed properties of galaxies, as already done using the DR1 data by \citet{Baes2024a, Baes2024b} and \citet{BokonaTulu2026}. This requires matched dust-free data products for each dust-aware\footnote{We use the term `dust-aware' to refer to data products based on radiative transfer calculations with dust attenuation and dust emission.} flux or image. We have therefore run a second set of {\texttt{SKIRT}} simulations in the {\tt{NoMedium}} simulation mode. For these simulations we used the same {\texttt{BPASS}} SSP templates for the stellar particles older than 10~Myr as also done for the standard, dust-aware simulations. For the star-forming gas cells, we adopted the dust-free {\texttt{TODDLERS}} templates generated by \citet{Kapoor2024}. These templates are fully consistent with the templates used in the dust-aware simulations. They do not consider attenuation and thermal emission by dust, but they account for attenuation by gas and the corresponding nebular emission lines. 

Dust-free simulations are computationally much cheaper than simulations that include dust attenuation and thermal dust emission. We could therefore afford to increase the number of photon packets by an order of magnitude, from $10^9$ to $10^{10}$, which reduces the Monte Carlo noise (see Sect.~{\ref{Uncertainties.sec}} for further details on Monte Carlo noise). The output of the dust-free simulations consists of similar data cubes as the dust-aware simulations. 

In addition to the synthetic dust-aware and dust-free observations, TSA DR2 provides an expanded set of intrinsic physical property maps generated through dedicated probes in {\texttt{SKIRT}}. As in DR1, we include maps of the stellar mass surface density, the stellar-mass-weighted mean stellar age, the stellar-mass-weighted mean stellar metallicity, and the dust mass surface density. DR2 introduces several additional probes: the SFR surface density\footnote{The SFRs in this work are `instantaneous' SFRs and are derived by summing the SFRs of the individual star-forming gas cells in the simulated galaxies, rather than from the stellar particles that formed over a finite time period.}, the interstellar gas surface density, and the ISM-gas-mass-weighted mean metallicity. These maps were calculated by projecting the intrinsic TNG50 particle and cell properties on the plane of the sky. Together, these probes enable detailed spatially resolved studies of the interplay between stars, gas, dust, and star formation across the full wavelength range covered by the atlas.

\begin{figure*}
\includegraphics[width=\textwidth]{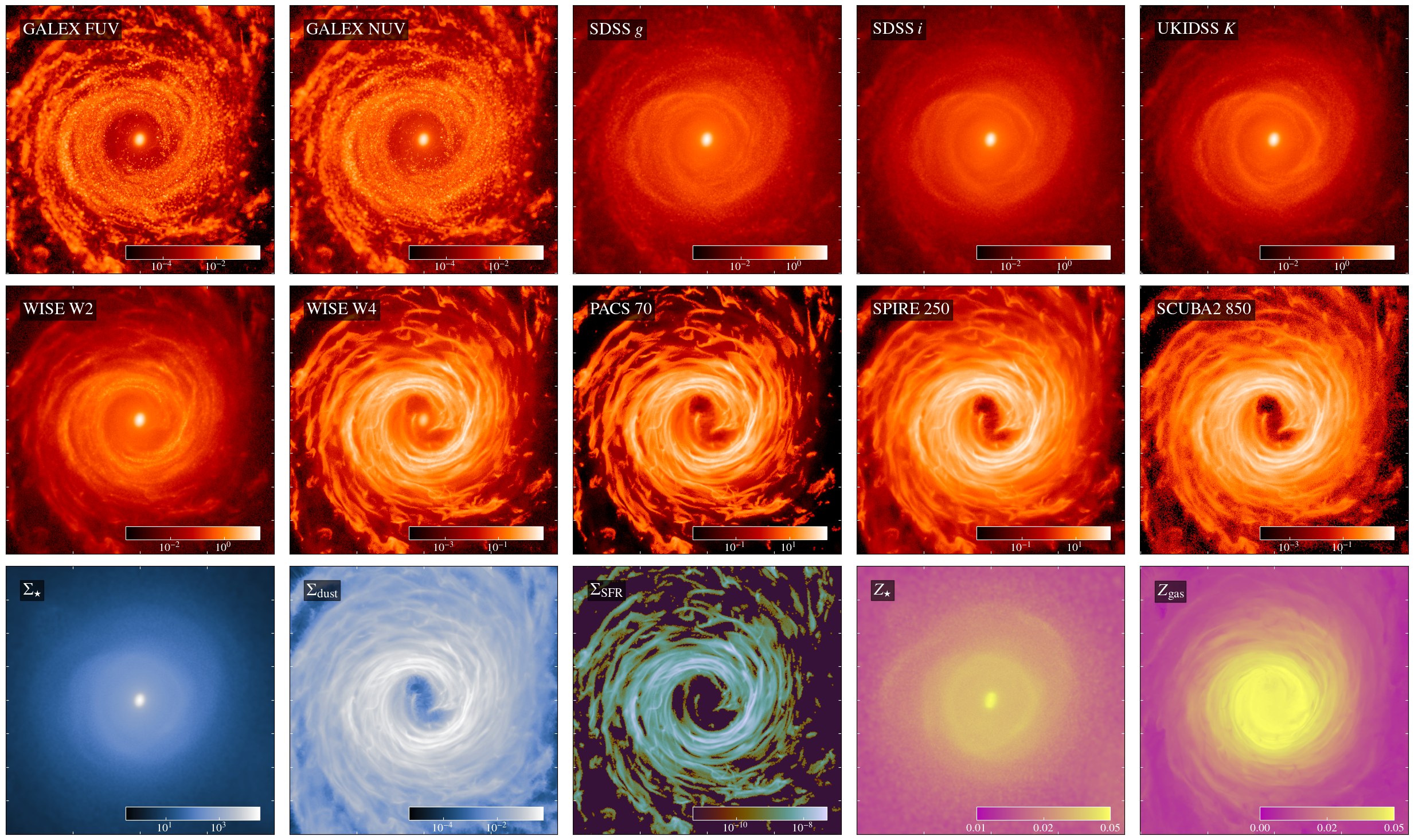}
\caption{Multi-wavelength synthetic images and physical property maps for the galaxy TNG\,294866, observed from viewing orientation O2, corresponding to an inclination of 18.9~deg. All images and maps have a field of view of 80 kpc $\times$ 80 kpc, and a pixel scale of 100 pc. The first and second rows show a representative selection of synthetic images, covering the UV to submm wavelength range. These bands are chosen to illustrate the diversity of emission mechanisms across the spectrum and do not represent the full set of available filters. All images are shown in surface-brightness units of MJy~sr$^{-1}$. The third row presents corresponding physical property maps extracted from the underlying simulation, including the stellar mass surface density (in units of M$_\odot$\,pc$^{-2}$), dust mass surface density (M$_\odot$\,pc$^{-2}$), SFR surface density (M$_\odot$\,yr$^{-1}$\,pc$^{-2}$), stellar metallicity (dimensionless), and gas-phase metallicity (dimensionless).}
\label{Images_TNG294866_O2.fig}
\end{figure*}

\section{Atlas presentation}
\label{Presentation.sec}

\subsection{Images and physical property maps}
\label{ImagesMaps.sec}

A central component of the TSA DR2 is the set of synthetic images and physical property maps generated with the Monte Carlo radiative transfer code {\texttt{SKIRT}}. For each galaxy in the atlas, we provide spatially resolved spectral cubes covering the UV to submm range, images in common broadband filters, and matching maps of key physical quantities. All data products are generated in a fully self-consistent manner, using the same underlying galaxy model and radiative transfer solution, thereby ensuring physical coherence across wavelengths.

The synthetic images are generated in a number of common broadband filters: we have selected \textit{GALEX}, SDSS {\em{ugriz}}, UKIDSS {\em{YJHK}}, \textit{WISE}, \textit{Herschel}, and SCUBA-2. For each galaxy, images are provided for multiple viewing orientations, allowing the effects of inclination and dust geometry on the emergent emission to be explored. The images are delivered in physical surface-brightness units (MJy sr$^{-1}$) and preserve the full spatial resolution of the radiative transfer calculation. As discussed in Sect.~{\ref{Probes.sec}}, we also provide physical property maps corresponding to the same observer positions as the images. These maps provide essential context for interpreting the multi-wavelength emission and facilitate direct connections between observed morphology and the underlying physical structure of the galaxies.

Figure~{\ref{Images_TNG294866_O2.fig}} illustrates the range of information available for a representative galaxy in the atlas. The panels on the first and second rows show synthetic images in selected broadband filters from the UV to the submm, highlighting the transition from clumpy, attenuation-dominated emission at short wavelengths to smoother dust emission at longer wavelengths. The bottom panels present the corresponding physical property maps, demonstrating how variations in stellar mass, SFR, dust content, and metallicity shape the observed SED and morphology. This figure exemplifies the internal consistency of the atlas products and the close coupling between emission properties and physical structure.

\begin{figure*}
\includegraphics[width=\textwidth]{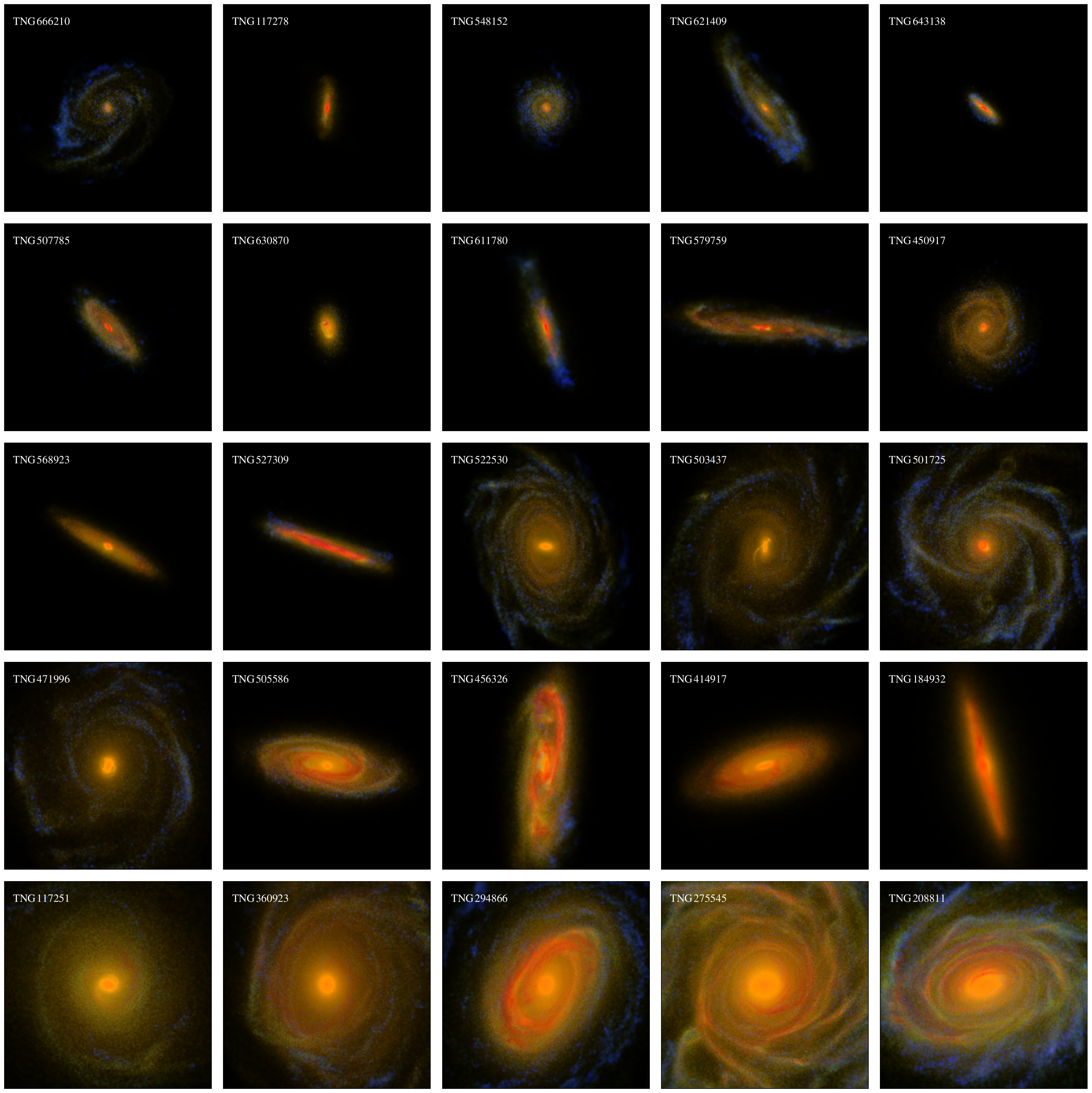}
\caption{RGB composite images for a representative subset of galaxies from the TSA DR2, constructed from synthetic {\em{GALEX}} FUV (blue), SDSS $g$ (green), and UKIDSS $J$ (red) images generated with {\texttt{SKIRT}}. The composites were generated using the algorithm of \citet{Lupton2004}, as implemented in the \texttt{Astropy} package \citep{AstropyCollaboration2013}. We adopted multiplicative scaling factors of 15, 3, and 1 for the FUV, $g$, and $J$ bands, respectively, together with the parameters $\texttt{stretch}=1$ and $\texttt{Q}=8$. All galaxies are shown from the O1 viewing orientation, corresponding to random lines of sight through the simulated systems. The galaxies are ordered by increasing stellar mass from left to right and top to bottom. Each panel shows a $60 \times 60$ kpc field of view. This zoomed-in region does not encompass the full spatial extent of the galaxies, thereby highlighting their internal structure.}
\label{Sample25_uvoir.fig}
\end{figure*}

\begin{figure*}
\includegraphics[width=\textwidth]{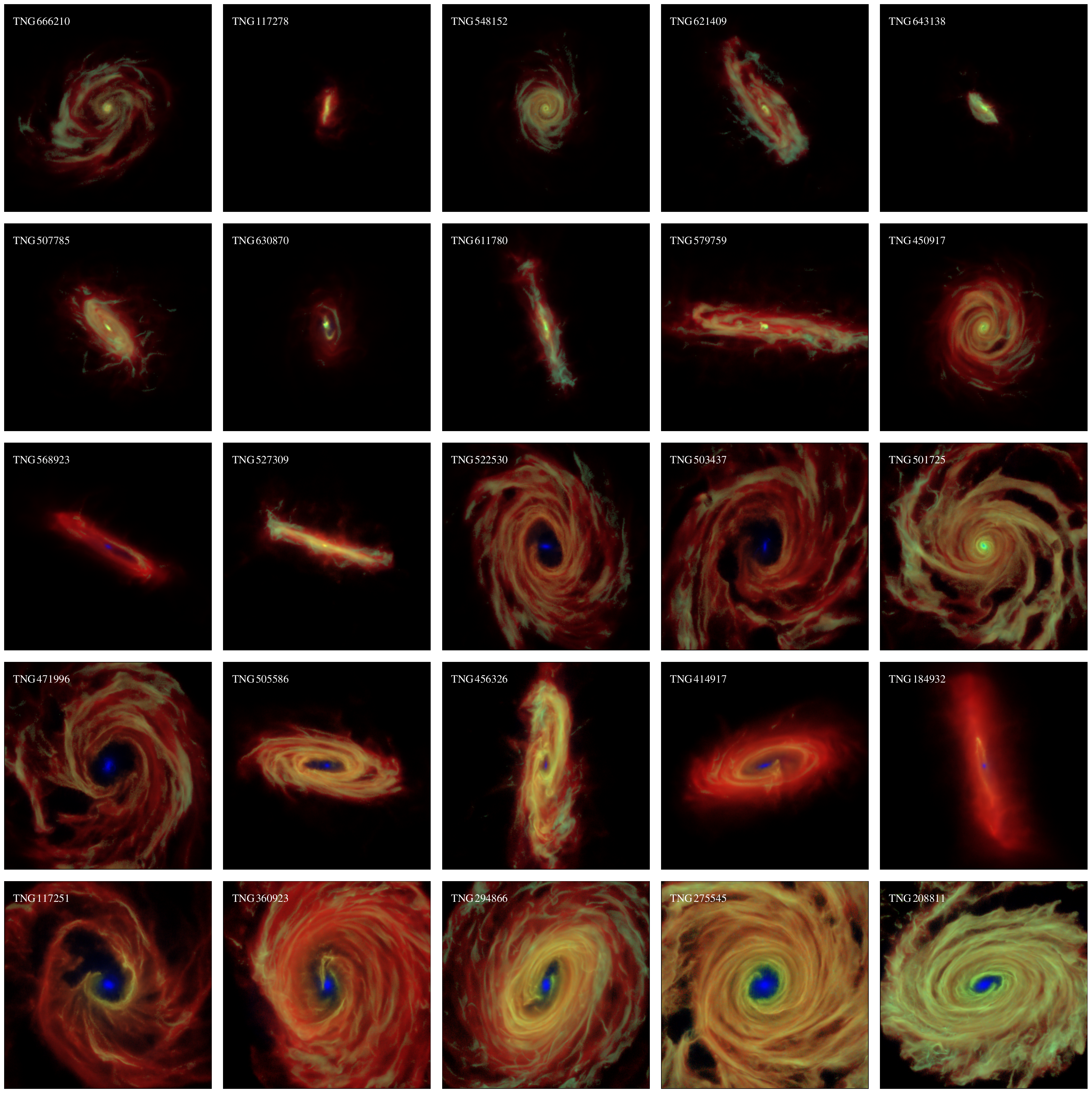}
\caption{RGB composite images for a representative subset of galaxies from the TSA DR2, constructed from synthetic WISE W4 (blue), PACS 160~$\mu$m (green), and SPIRE 350~$\mu$m (red) images generated with {\texttt{SKIRT}}. The composites were generated using the algorithm of \citet{Lupton2004}, as implemented in the \texttt{Astropy} package \citep{AstropyCollaboration2013}. We adopted multiplicative scaling factors of 40, 1, and 4 for the W4, 160~$\mu$m, and 350~$\mu$m bands, respectively, together with the parameters $\texttt{stretch}=10$ and $\texttt{Q}=15$. All galaxies are shown from the O1 viewing orientation, corresponding to random lines of sight through the simulated systems. The galaxies are ordered by increasing stellar mass from left to right and top to bottom. Each panel shows a $60 \times 60$ kpc field of view. This zoomed-in region does not encompass the full spatial extent of the galaxies, thereby highlighting the spatial distribution of the dust emission.}
\label{Sample25_irsubmm.fig}
\end{figure*}

To illustrate the diversity of galaxy morphologies and dust geometries in the TSA DR2, we present in Figs.~{\ref{Sample25_uvoir.fig}} and~{\ref{Sample25_irsubmm.fig}} two montage figures composed of RGB composite images for a representative subset of galaxies. These galaxies correspond to the carefully selected sample used in recent studies by \citet{EuclidCollaborationAbdurrouf2025} and \citet{EuclidCollaborationNersesian2026} to cover the ($M_\star$, sSFR)--plane (see Sect. 3.1 of the former paper). The first montage (Fig.~{\ref{Sample25_uvoir.fig}}) combines UV, optical, and NIR bands, highlighting unobscured star formation, stellar populations, and dust attenuation features. The second montage (Fig.~{\ref{Sample25_irsubmm.fig}}) combines MIR, FIR, and submm bands, emphasising the distribution of dust emission and its relation to the underlying galaxy structure. Together, these figures provide a visual overview of the breadth of galaxy types included in the atlas and demonstrate the capability of the TSA DR2 to capture the multi-wavelength appearance of galaxies in a physically consistent radiative transfer framework.

\subsection{Integrated galaxy properties}
\label{IntegratedProperties.sec}

\begin{figure}
\centering
\includegraphics[width=\columnwidth]{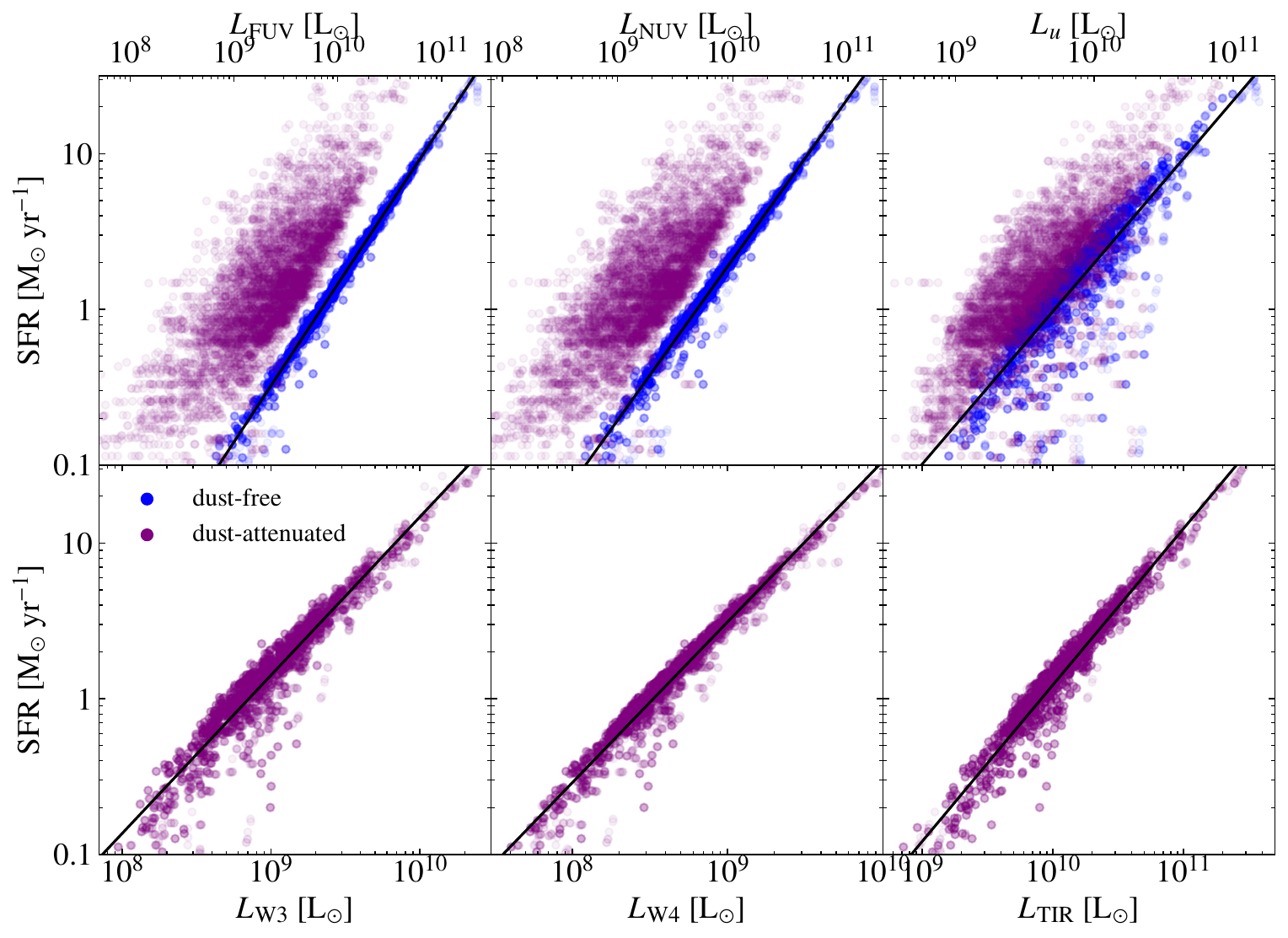}%
\caption{Correlation between integrated luminosities in selected broadband filters and the SFR for TSA DR2 galaxies. The sample includes all galaxies observed in the four independent random viewing orientations O1--O4.  The top row shows the {\em{GALEX}} FUV and NUV bands and the SDSS {\em u} band, while the bottom row shows the {\it{WISE}} W3 and W4 bands and the total infrared luminosity. For the UV and optical bands, both dust-free luminosities (blue symbols) and dust-attenuated luminosities (purple symbols) are shown, illustrating the impact of dust attenuation on the observed relations. For the infrared bands, only dust-aware luminosities are shown, as these bands trace dust emission. Solid lines indicate linear fits presented in Eq.~(\ref{SFRfits}), fitted to the dust-free luminosities for the UV and optical bands and to the dust-aware luminosities for the infrared bands. The corresponding fit parameters are listed in Table~\ref{tab:sfr_fits}.}
\label{fig:SFR_estimators}
\end{figure}

In addition to the spatially resolved images and physical property maps presented in Sect.~{\ref{ImagesMaps.sec}}, TSA DR2 provides a table of integrated galaxy properties derived from these maps. For each galaxy in the atlas and for each of the seven viewing orientations, we report a set of global physical quantities together with integrated photometric measurements. This table is intended to support analyses based on unresolved galaxy properties and to enable direct comparisons with observational studies that rely on integrated photometry.

The physical quantities included in the catalogue are obtained by integrating the corresponding physical property maps over the full 80 kpc $\times$ 80 kpc field of view. These include stellar mass, ISM gas mass, dust mass, SFR, specific SFR (sSFR), mean stellar age, and mean stellar and gas-phase metallicity. The catalogue also provides integrated absolute magnitudes in the AB system for a broad set of commonly used photometric bands, spanning the UV to the submm range. These magnitudes are measured directly from the synthetic images generated with {\texttt{SKIRT}}. For the UV to NIR bands, both dust-free and dust-aware magnitudes are provided, allowing users to explicitly assess the impact of dust attenuation and dust emission on integrated galaxy observables.

\begin{table*}
\caption{Power-law fits between integrated luminosities and the reference SFR for TSA DR2 galaxies, following Eq.~(\ref{SFRfits}).}
\label{tab:sfr_fits}
\centering
\begin{tabular}{lcccc}
\hline\hline\\[-0.7ex]
band & slope $a$ & intercept $b$ & $R^2$ & $r$ \\[1.5ex]
\hline
\\[-0.7ex]
FUV (dust-free) & $1.014 \pm 0.001$ & $-9.98 \pm 0.01$ & 0.988 & 0.994\\
NUV (dust-free) & $1.035 \pm 0.002$ & $-10.03 \pm 0.02$ & 0.980 & 0.990 \\
{\em{u}} (dust-free) & $1.042 \pm 0.006$ & $-10.12 \pm 0.06$ & 0.806 & 0.898 \\
W3       & $1.017 \pm 0.004$ & $-9.00 \pm0.04$ & 0.912 & 0.955 \\
W4       & $1.032 \pm 0.003$ & $-8.80 \pm 0.02$ & 0.961 & 0.980 \\
TIR       & $1.004 \pm 0.003$ & $-9.95 \pm 0.03$  & 0.945 & 0.972 \\[1.5ex]
\hline
\end{tabular}
\end{table*}

As a sanity check on the radiative transfer simulations and a demonstration of the potential of the TSA DR2, Fig.~{\ref{fig:SFR_estimators}} compares the integrated SFR of all galaxies with ${\text{SFR}} > 0.1~{\text{M}}_\odot\,{\text{yr}}^{-1}$ to the luminosities in a number of broadband filters. We use galaxies in the random orientations O1--O4; the fifth random orientation O5 is antipodal to O4 and therefore does not provide independent information \citep[][Sect.~2.4.3]{Baes2024a}. The first five panels show luminosities in the {\em{GALEX}} FUV and NUV bands, the SDSS {\em{u}} bands, and the {\it{WISE}} W3 and W4 bands. For these broadband filters, the luminosity is defined as $\nu L_\nu$, with $\nu$ the effective frequency of the band. The last panel shows the integrated total infrared luminosity $L_{\text{TIR}}$, calculated using the five-band formula of \citet{Galametz2013},
\begin{multline}
L_{\text{TIR}} = 2.023\,L_{24} + 0.523\,L_{70} \\
+ 0.390\,L_{100} + 0.577\,L_{160} + 0.721\,L_{250}.
\end{multline}
For the UV/optical bands we show both dust-attenuated luminosities (pink) and dust-free luminosities (blue), whereas the bottom panels only show dust-aware luminosities, since these bands trace dust in emission. The solid lines are linear fits of the form
\begin{equation}
\log\left( \frac{{\text{SFR}}}{{\text{M}}_\odot\,{\text{yr}}^{-1}} \right)
=
a \log \left( \frac{L_{\text{band}}}{{\text{L}}_\odot} \right) + b,
\label{SFRfits}
\end{equation}
In the upper row panels, the line is fit to the dust-free rather than the dust-attenuated luminosities. The slopes, intercepts, $R^2$ values, and Pearson correlation coefficients, $r$, of these fits are listed in Table~{\ref{tab:sfr_fits}}.

UV continuum primarily traces young massive stars, sensitive to timescales of the order of 100 Myr. Our dust-free luminosities exhibit tight, approximately linear correlations with the reference SFR, demonstrating that these bands are intrinsically strong tracers of recent star formation when attenuation effects are absent. Because the reference SFR is effectively instantaneous, some scatter is expected for UV tracers that respond to $\sim$100 Myr timescales. Dust attenuation strongly affects UV radiation, and as a result, the dust-attenuated UV luminosities show substantial scatter and systematic deviations. The estimation of SFRs from observed dust-attenuated {\em{GALEX}} luminosities therefore requires a dust attenuation correction \citep[e.g.,][]{Salim2007, Davies2016, Brown2017}. 

Emission in the SDSS {\em{u}} band is sensitive to recent star formation but has non-negligible contamination by old stellar populations, making it a less reliable SFR estimator. We find a nearly linear relation between the SFR and the unattenuated {\em{u}}-band luminosity, with a significant scatter that increases toward low-SFR systems, where it is expected that the older stellar populations contribute a relatively larger fraction \citep{Bell2003, Hopkins2003, Zhou2017}. On the other hand, attenuation by dust is less severe than in the {\em{GALEX}} UV bands, which can clearly be seen in the top right panel of Fig.~{\ref{fig:SFR_estimators}}. 

The MIR is typically dominated by dust emission from star-forming regions, with a smaller contribution from stellar continuum, making them popular tracers for the SFR \citep[e.g.,][]{Calzetti2007, Calzetti2010, Zhu2008, Cluver2017, Cluver2025}. Galaxy-dependent differences in the ratio of obscured to unobscured star formation, dust heating by old stellar populations, and metallicity and PAH effects often lead to a sublinear relation between SFR and MIR luminosity \citep{Calzetti2010, Figueira2022, Cluver2025}. For our TSA galaxies, we find a nearly linear relation between the SFR and the monochromatic luminosity in both the {\it{WISE}} W3 and W4 bands (bottom-left and bottom-central panels of Fig.~{\ref{fig:SFR_estimators}}). The scatter is slightly larger for the W3 band. The near-linearity found for TSA galaxies likely reflects both the controlled physical conditions of the simulated sample and the absence of heterogeneous observational systematics present in real surveys. We note that the slopes found for the {\it{WISE}} bands agree very well with the results from \citet{Boquien2021}.

The total infrared luminosity (bottom-right panel of Fig.~{\ref{fig:SFR_estimators}}), a direct measure for the bolometric re-emission by interstellar dust, is often considered one of the most reliable tracers of the star formation obscured by dust \citep[e.g.,][]{Kennicutt1998, Kennicutt2009}. For our TSA galaxies, we find an almost one-to-one correspondence between $L_{\text{TIR}}$ and SFR.

Overall, this analysis demonstrates that the integrated luminosities provided in TSA DR2 exhibit physically sensible and internally consistent correlations with the underlying star formation activity of the simulated galaxies. The fitted relations shown in Fig.~\ref{fig:SFR_estimators} are not immediately intended as new calibrations, but rather as a diagnostic of the internal consistency of the atlas products. A more detailed analysis of SFR tracers based on the TSA DR2, both on global and on sub-kpc scales, will be presented in forthcoming work.

\subsection{Data access and availability}

The TSA DR2 data products are made available through a combination of direct download and on-demand access. The catalogue of integrated galaxy properties and the set of broadband images (total volume $\sim$811~GB) are distributed via the IllustrisTNG Public Data Access server.\footnote{\url{https://www.tng-project.org/data/}}

The full set of spatially resolved spectral data cubes amounts to a total volume of approximately 14~TB. Due to their size, these data products are not distributed as a single bulk download. Instead, the data cubes are available upon request and are typically provided on a per-galaxy basis. For larger data volumes, alternative transfer methods may be arranged. We are exploring options to provide more flexible access to these data products, including remote querying and analysis services.

\section{Comparison between DR1 and DR2}
\label{DR12.sec}

\subsection{SEDs and images}

\begin{figure*}
\centering
\includegraphics[width=0.47\textwidth]{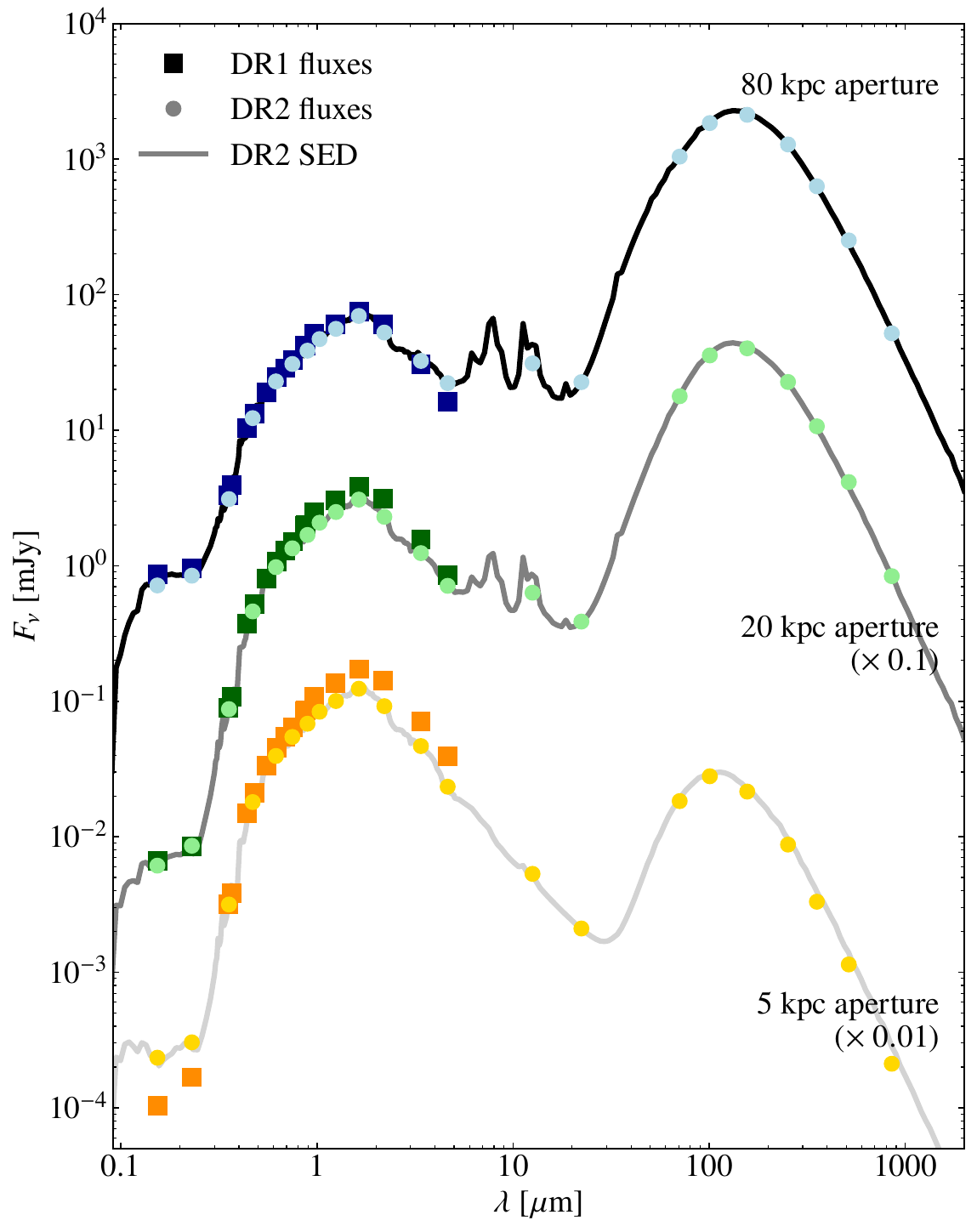}
\hspace*{2em}
\includegraphics[width=0.47\textwidth]{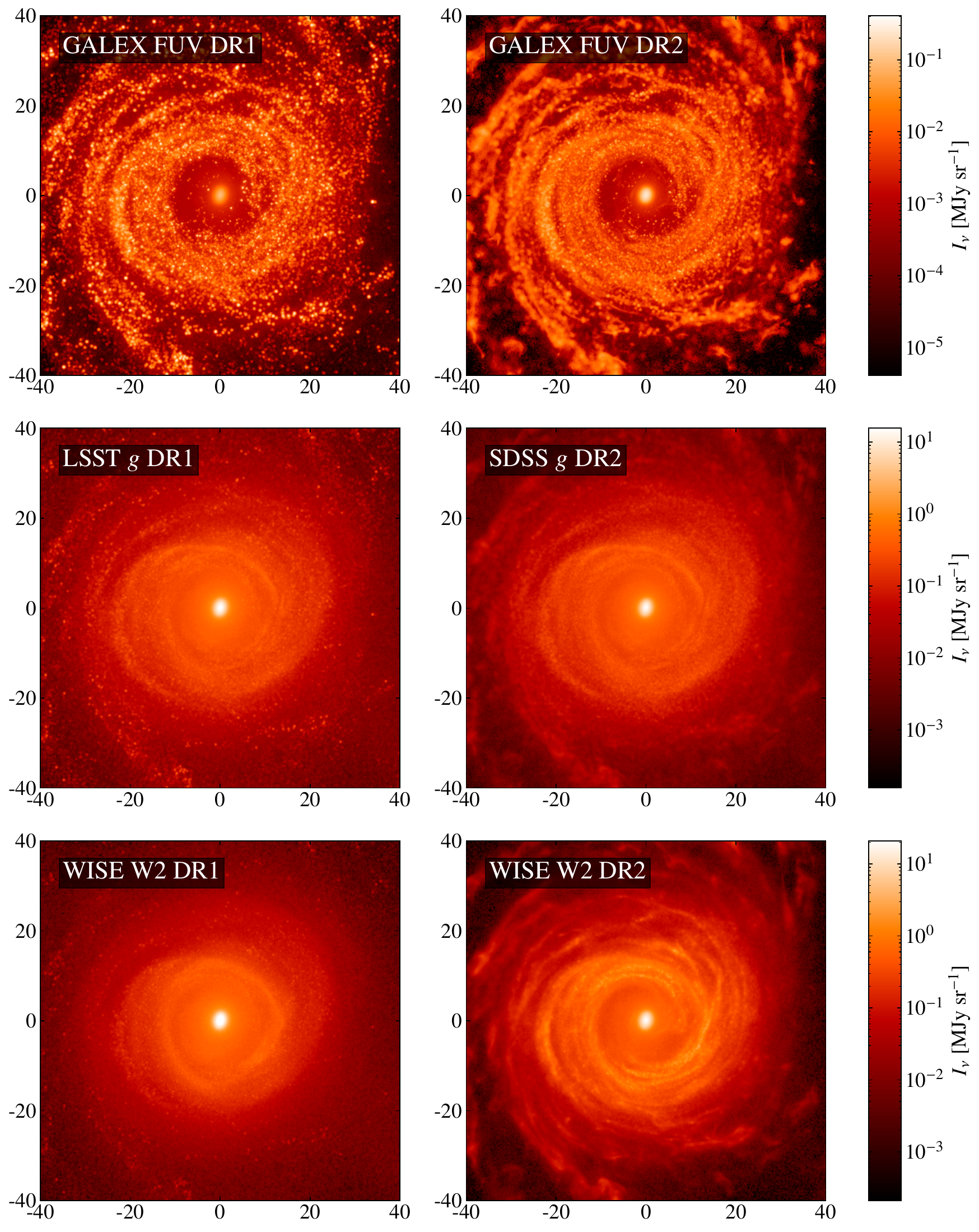}
\caption{Comparison between TSA DR1 and DR2 for the galaxy TNG\,294866, observed from viewing orientation O2. 
\emph{Left:} Integrated SEDs measured within three square apertures of side length 5~kpc, 20~kpc, and 80~kpc. DR1 fluxes are available only in a limited set of UV to NIR bands, while DR2 provides a continuous SED together with broadband fluxes extending from the UV to the submm. The SEDs are scaled for clarity as indicated in parentheses. In addition to differences in total flux, the SED shape varies with aperture, with the smallest aperture showing a markedly reduced FIR to optical ratio due to a central deficit in the dust distribution (see Fig.~\ref{Images_TNG294866_O2.fig}). 
\emph{Right:} Comparison of selected synthetic images from TSA DR1 and DR2 in representative bands, illustrating the overall consistency in spatial structure and the differences arising from updated stellar population and dust emission modelling in DR2.}
\label{Comparison_TNG294866_O2.fig}
\end{figure*}

An important improvement in TSA DR2 over the first data release is the availability of continuous SEDs and broadband fluxes extending from the UV to the submm. To illustrate both the continuity between the two releases and the additional information provided by DR2, Fig.~\ref{Comparison_TNG294866_O2.fig} presents a comparison of SEDs and selected images for the galaxy TNG\,294866, observed from viewing orientation O2. This is the same system shown in the multi-wavelength images and physical property maps in Fig.~\ref{Images_TNG294866_O2.fig}.

The left-hand panel of Fig.~\ref{Comparison_TNG294866_O2.fig} shows integrated SEDs measured within three square apertures of increasing size: 5~kpc, 20~kpc, and 80~kpc. The largest aperture corresponds to the full field of view of the DR2 images, while the smaller apertures probe increasingly central regions of the galaxy. For TSA DR1, only fluxes in a limited set of predefined photometric bands from the UV to the NIR are available. In contrast, TSA DR2 provides a densely sampled SED together with broadband fluxes covering the entire UV to submm range.

As expected, the total flux decreases with decreasing aperture size. However, the SED shape also varies systematically with aperture. In particular, the FIR-to-optical ratio is significantly lower for the smallest aperture than for the larger apertures. This behaviour directly reflects the spatial distribution of dust in this galaxy, which exhibits a central deficit in the dust mass surface density, as shown in Fig.~\ref{Images_TNG294866_O2.fig}. As a result, the central regions contribute relatively little FIR emission compared to stellar light, while dust emission becomes increasingly important at larger galactocentric radii. This example illustrates that spatially resolved information is essential not only for measuring total luminosities, but also for understanding how SED shapes depend on aperture and physical structure.

The right-hand panels of Fig.~\ref{Comparison_TNG294866_O2.fig} compare selected synthetic images from TSA DR1 and DR2 in representative bands. Overall, the spatial distribution of the emission is consistent between the two releases, while differences in colour and intensity reflect updates in the underlying stellar population and dust emission modelling. In DR2, the SEDs of older stellar populations are based on {\texttt{BPASS}} models rather than the BC03 models used in DR1, leading to modestly bluer UV and optical colours for evolved populations. The treatment of young stellar populations has also been updated, with DR2 adopting the {\texttt{TODDLERS}} templates instead of the {\texttt{MAPPINGS}}~III models used in DR1, together with a different sampling of young stellar particles. These changes primarily affect the UV and optical emission associated with recent star formation. The {\textit{WISE}} W2 images also differ between the two releases since DR1 does not contain dust emission, and some dust emission is noticeable in the DR2 image in this band. Interestingly, galaxy TNG\,294866 has a central hole in its ISM density distribution, a feature that is present in many TNG50 galaxies \citep[see e.g.][]{Gebek2023}. As a result, the contribution of dust emission to the {\textit{WISE}} W2 band is negligible in the central region, but it is clearly present in the image at larger radii and in the global SED.

The impact of these modelling updates is more pronounced for the smaller apertures. For the largest aperture, which averages over the entire galaxy, the DR1 and DR2 SEDs agree more closely, reflecting the reduced sensitivity to local modelling choices when integrating over the full system. This comparison demonstrates both the overall consistency between the two data releases and the increased flexibility and physical realism provided by TSA DR2, particularly for spatially resolved and aperture-dependent studies.

\begin{figure*}
\centering
\includegraphics[height=0.305\textwidth]{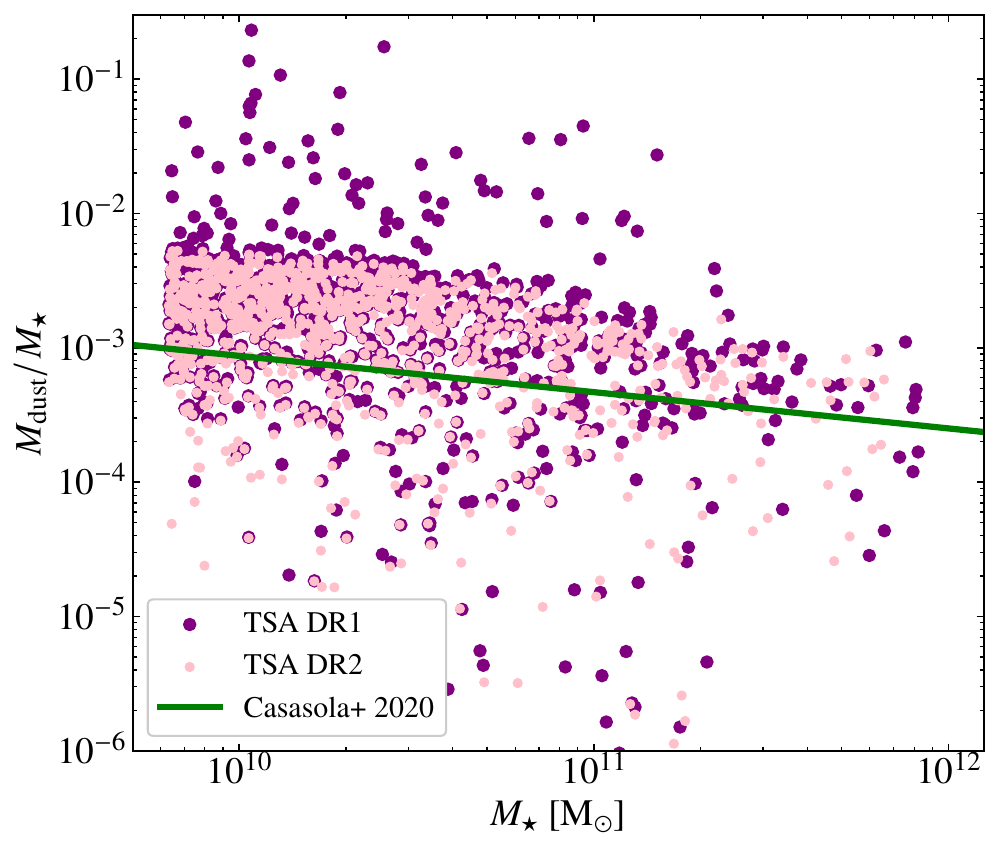}%
\quad
\includegraphics[height=0.305\textwidth]{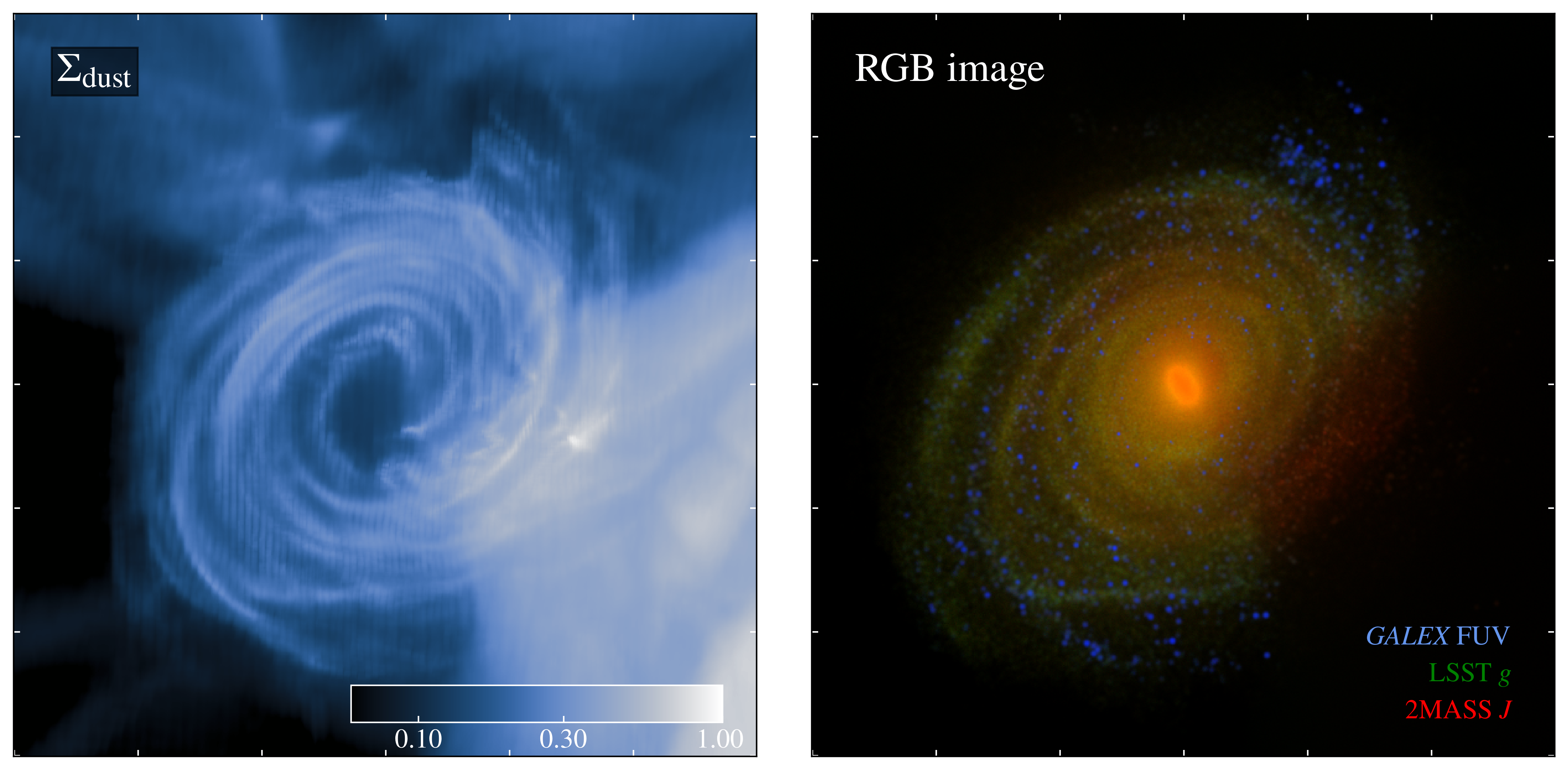}%
\caption{Illustration of dust-allocation artefacts identified in TSA DR1 and resolved in TSA DR2 (see Sect.~{\ref{DustAllocation.sec}}).
\emph{Left:} Stellar mass versus specific dust mass for TSA DR1 (dark symbols) and TSA DR2 (light symbols), compared to the average relation for late-type galaxies in the DustPedia sample by \citet{Casasola2020}. A subset of TSA DR1 galaxies exhibits anomalously high dust-to-stellar mass ratios, which are no longer present in TSA DR2.
\emph{Middle:} Dust mass surface density map for TNG\,143886 in TSA DR1. The galaxy is shown as seen from viewing position O1, and the field of view is 30 kpc $\times$ 30 kpc. An extended region of unrealistically high dust column density caused by the dust allocation artefact is notable in the bottom-right quadrant.
\emph{Right:} Corresponding optical three-colour image of the same galaxy, combining the {\em{GALEX}} FUV (blue), LSST {\em{g}} (green), and 2MASS {\em{J}} (red) bands. The spurious dust structure generates an excessive reddening in the bottom-right quadrant.}
\label{Anomalies.fig}
\end{figure*}

\subsection{Improvements of the dust allocation scheme}
\label{DustAllocation.sec}

The transition from TSA DR1 to DR2 represents not only an extension of the atlas in terms of wavelength coverage, but also an improvement in the robustness of the dust modelling. In the course of post-release validation of TSA DR1, a number of previously unidentified artefacts were discovered that affected a small but non-negligible subset of galaxies. These issues have been analysed in detail and resolved in TSA DR2.

The problem was first identified through visual analytics using the {\texttt{ARGOS}}\footnote{\url{https://skirt-argos.ugent.be/}} framework \citep{Vauterin2026}, which combines interactive data exploration with linked multi-dimensional visualisations. By inspecting correlations between integrated galaxy properties, a group of galaxies was found to deviate strongly from the main trends in the diagnostic relation between stellar mass and specific dust mass. The vast majority of galaxies populate a region in the $(M_\star, M_{\text{dust}}/M_\star)$--plane that is in good agreement with the locus occupied by observed galaxies \citep{Cortese2012, Smith2012, Casasola2020}. A subset of about 40 TSA DR1 galaxies, however,  occupy a region in this plane well above the location of the bulk of the galaxies, exhibiting unrealistically high dust masses for their stellar mass (see left panel of Fig.~{\ref{Anomalies.fig}}).

Subsequent visual inspection of the synthetic images and dust surface density maps for these galaxies revealed clear morphological anomalies. The middle and right panels of Fig.~{\ref{Anomalies.fig}} illustrate this for TNG\,143886, one of the galaxies with an excessive specific dust mass. The dust mass surface density map shows an extended region of unrealistically high dust column density, resulting in excessive reddening in optical images. Importantly, the corresponding gas and stellar distributions in the original TNG50 simulation data appear normal, indicating that the artefacts were introduced during the {\texttt{SKIRT}} radiative transfer post-processing rather than originating from the hydrodynamical simulation itself.

A detailed investigation of the three-dimensional dust distribution traced the origin of these artefacts to the dust allocation scheme used in TSA DR1. In that release, the dust density in each cell of the radiative transfer grid was derived from the local gas density, as described in Sect.~2.4.2 of \citet{Baes2024a}. However, the gas distribution imported into {\texttt{SKIRT}} is represented on a Voronoi mesh that is subject to culling and re-gridding during the construction of the radiative transfer grid. In particular, cells associated with circumgalactic gas are discarded because they are assumed to be dust-free, and only the cells corresponding to the interstellar medium are retained \citep{Torrey2012, Torrey2019}. As a result, cells located near the interface between retained and discarded regions can become artificially large.

When dust densities are assigned based on gas density rather than gas mass, such changes in cell volume can lead to severe overestimates of the dust mass in affected cells. In extreme cases, a single enlarged cell can dominate the total dust mass of the galaxy and produce unphysical attenuation and emission signatures. This behaviour was confirmed through targeted inspection of affected galaxies and analysis of the corresponding grid construction and dust assignment steps.

In TSA DR2, this issue has been resolved by modifying the dust allocation scheme to ensure mass conservation. Dust is now assigned based on the gas mass rather than the gas density in each cell, eliminating the sensitivity to changes in cell volume introduced during grid construction. This approach preserves the total dust mass by construction and prevents the formation of spurious high-density dust regions at grid boundaries. All galaxies in TSA DR2 were fully reprocessed using the revised dust allocation scheme. Extensive validation has confirmed that the anomalous galaxies identified in TSA DR1 no longer exhibit unphysical dust masses or attenuation features in TSA DR2. The effect is illustrated in the left panel of Fig.~{\ref{Anomalies.fig}}.

An alternative way to avoid this class of interpolation artefacts would have been to perform the radiative transfer calculations directly on the native Voronoi mesh of the TNG50 simulation. As discussed in Sect.~{\ref{DustyMedium.sec}}, we chose to use an adaptive octree grid because it offers substantially better computational performance for the radiative transfer calculations, while there is no compelling reason to regard the native Voronoi mesh as the optimal discretisation for radiative transfer. With the revised dust assignment procedure described above, the interpolation artefacts introduced by the regridding are eliminated, allowing us to retain the advantages of the adaptive octree without compromising the physical reliability of the resulting dust distributions.

This episode highlights the importance of complementary quality-control approaches in large, complex data products. While standard numerical diagnostics did not flag the issue, hypothesis-generating visual exploration proved essential for identifying subtle but astrophysically significant artefacts. The resulting improvements in the dust allocation scheme significantly enhance the reliability of the TSA DR2 products and underscore the value of iterative validation combining quantitative analysis with detailed visual inspection.

\section{Monte Carlo uncertainties and reliability}
\label{Uncertainties.sec}

\subsection{Pixel-level uncertainty assessment}

\begin{figure*}
\includegraphics[width=\textwidth]{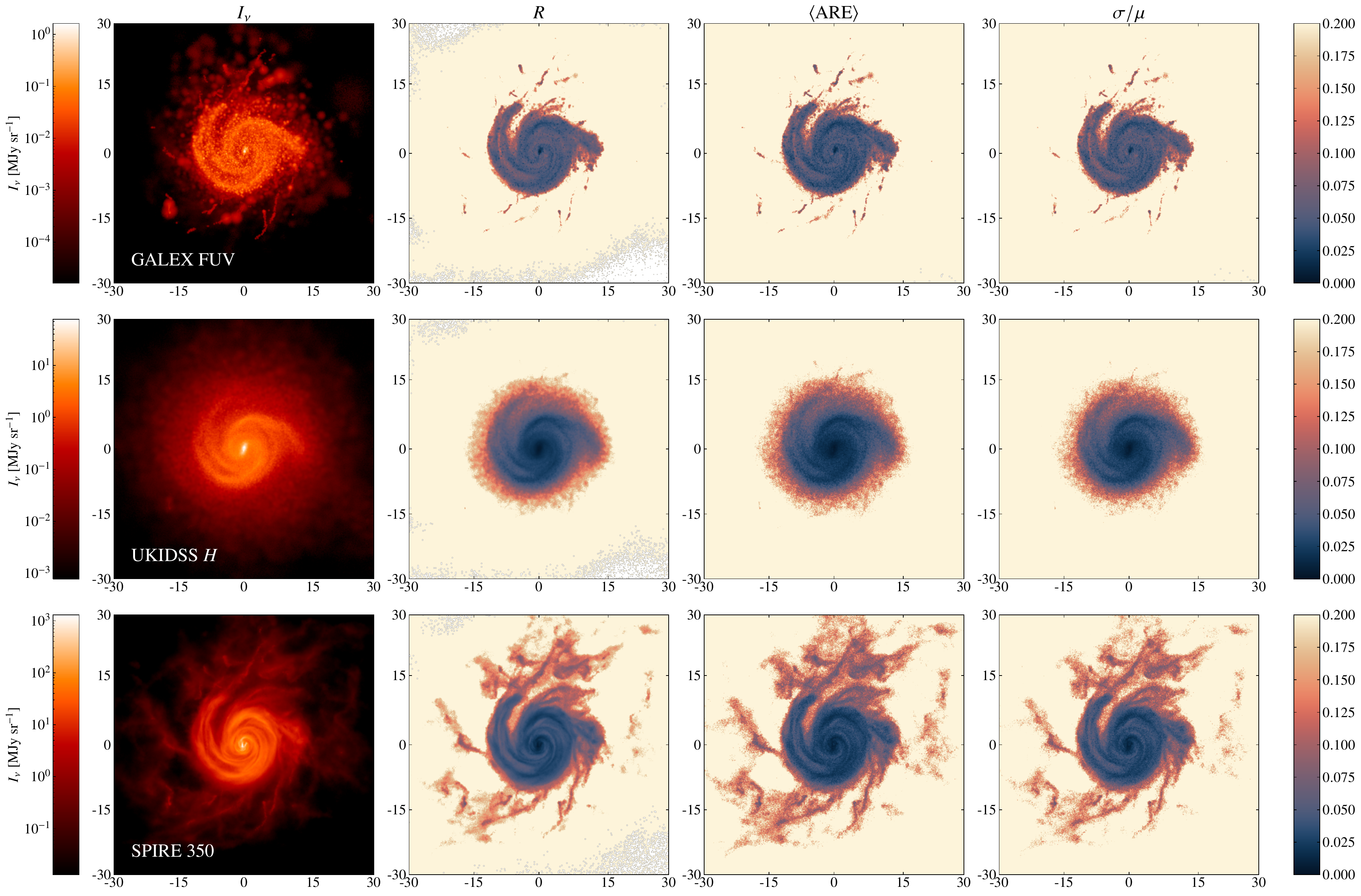}
\caption{Assessment of Monte Carlo noise in synthetic images generated with {\texttt{SKIRT}}, illustrated for the galaxy TNG000008 observed in the face-on orientation.
Results are shown for three representative bands spanning the UV, near-infrared, and far-infrared regimes.
For each band, the relative Monte Carlo uncertainty is estimated using three independent approaches: 
(i) the reliability statistic $R$ computed internally by {\texttt{SKIRT}}; 
(ii) the absolute relative error between a reference high–signal-to-noise image and multiple lower–signal-to-noise realisations with different random seeds; 
and (iii) the pixel-wise signal-to-noise ratio derived from the mean and variance of multiple independent Monte Carlo realisations.}
\label{Uncertainties_TNG000008_FO.fig}
\end{figure*}

The synthetic images in the TSA DR2 are generated using the Monte Carlo radiative transfer code {\texttt{SKIRT}}. As a result, individual pixel values are subject to stochastic noise arising from the finite number of photon packets used to sample the radiative transfer process. Quantifying these uncertainties is important for assessing the reliability of pixel-level measurements, in particular for spatially resolved analyses and for studies targeting low surface brightness regimes. While early work has demonstrated the importance of uncertainties in synthetic Monte Carlo imaging\citep[e.g.,][]{Gordon2001}, this is seldom characterised explicitly in postprocessing work. Providing such guidance is therefore an important component of the TSA DR2.

{\texttt{SKIRT}} includes built-in statistical diagnostics to estimate the relative uncertainty due to Monte Carlo sampling, based on higher-order moments of the photon packet weight distribution accumulated in each detector bin. These diagnostics are described in detail in \citet{Camps2018b, Camps2020}, and are based on techniques used in Monte Carlo neutron transport \citep{X5MonteCarloTeam2003}. These diagnostics yield a dimensionless reliability parameter $R$, which provides an estimate of the relative Monte Carlo error for linear observables. Values of $R \lesssim 0.1$ are commonly taken as indicative of convergence, although this threshold should be regarded as approximate and context dependent.

While the interpretation of $R$ is straightforward for primary emission, uncertainty estimation in infrared bands is more complex. In these bands, the observed emission arises from dust grains heated by the local radiation field, such that uncertainties in the estimated radiation field propagate non-linearly into the dust emission spectrum. To assess whether Monte Carlo noise nevertheless remains the dominant source of pixel-scale uncertainty in this regime, we employed three complementary approaches. First, we analysed the internal {\texttt{SKIRT}} reliability parameter $R$ for each image. Second, we generated reference images using new {\texttt{SKIRT}} simulations with ten times more photon packets and treated these as effective high signal-to-noise realisations, computing the absolute relative error between the standard and high signal-to-noise images. This comparison was repeated for ten independent realisations of the standard simulation using different random number generators, yielding an average absolute relative error. Third, we generated ten independent image sets with identical input parameters but different random number generators and estimated the signal-to-noise ratio as the ratio of the mean to the standard deviation of the resulting pixel intensities.

For the test case shown in Fig.~{\ref{Uncertainties_TNG000008_FO.fig}} (galaxy TNG\,000008 in a face-on orientation), the three methods yield broadly consistent uncertainty estimates across UV, NIR, and submm bands. This agreement indicates that, for these images, the dominant contribution to the pixel-scale variance is indeed the stochastic noise associated with Monte Carlo sampling, even in bands dominated by secondary dust emission. In particular, no evidence is found for an additional source of variance arising from the non-linear propagation of radiation-field uncertainties into the dust emission that would exceed the Monte Carlo noise captured by the reliability statistics.

Several caveats are nevertheless important to emphasize. First, the diagnostics discussed here primarily quantify stochastic variance and do not directly address potential biases. Second, the limited sampling of the primary light source, namely the stellar component of the cosmological simulation, constitutes an independent source of uncertainty. In low surface brightness regions, the finite number of stellar particles and the interpolation schemes used to construct a continuous emissivity distribution can introduce fluctuations and systematic effects that are not captured by Monte Carlo radiative transfer diagnostics alone. Similarly, the discretisation of the dust distribution forms an additional systematic uncertainty. The uncertainty estimates presented here therefore quantify the Monte Carlo component of the error budget, but do not account for uncertainties associated with sparse stellar sampling or with the underlying physical modelling assumptions.

\subsection{Practical guidance}

\begin{figure}
\includegraphics[width=\columnwidth]{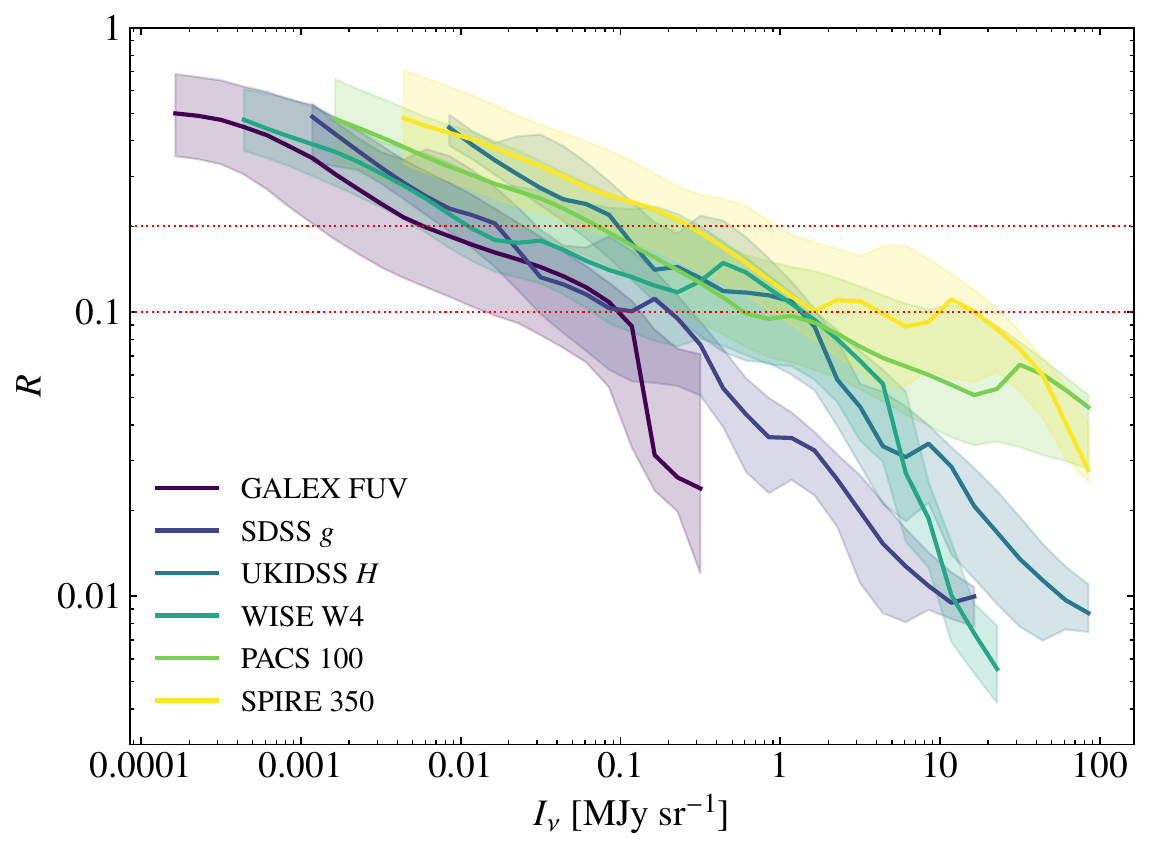}
\caption{Monte Carlo reliability of the TSA DR2 synthetic images as a function of surface brightness. For each photometric band, the distribution of the {\texttt{SKIRT}} reliability statistic $R$ is evaluated in bins of surface brightness using a representative sample of galaxies and viewing orientations. Solid lines indicate the median $R$ in each bin, while the shaded regions show the interquartile range, characterising the typical scatter at fixed surface brightness. Horizontal reference lines mark the commonly used reliability thresholds $R = 0.1$ and $R=0.2$.}
\label{RversusInu.fig}
\end{figure}

To provide practical guidance for users of the TSA DR2, we quantify how the Monte Carlo reliability parameter $R$ varies as a function of surface brightness for a representative subset of the atlas. For the same sample of 25 galaxies shown in Sect.~{\ref{ImagesMaps.sec}}, we computed $R$ maps in all available bands and combined the results across galaxies and viewing orientations. For each band, we evaluated the distribution of $R$ in bins of surface brightness and derived the median as well as the interquartile range, thereby characterising both the typical behaviour and the scatter of the Monte Carlo uncertainty as a function of surface brightness. The results are shown in Fig.~{\ref{RversusInu.fig}} for a number of representative broadbands. 

\begin{table}
\caption{Intensity percentiles for each photometric band, in MJy sr$^{-1}$.}
\label{tab:intensity_percentiles}
\centering
\begin{tabular}{lccc}
\hline\hline\\[-0.7ex]
Filter & $I_{50}$ & $I_{75}$ & $I_{90}$ \\[1.5ex]
\hline
\\[-0.7ex]
{\em{GALEX}} FUV   & 0.0615 & 0.0996 & 0.130 \\
{\em{GALEX}} NUV   & 0.0474 & 0.0768 & 0.100 \\
SDSS {\em{u}}      & 0.118 & 0.164 & 0.205 \\
SDSS {\em{g}}      & 0.0259 & 0.191 & 0.266 \\
SDSS {\em{r}}     & 0.0565 & 0.328 & 0.459 \\
SDSS {\em{i}}      & 0.0812 & 0.429 & 0.613 \\
SDSS {\em{z}}      & 0.0885 & 0.469 & 0.716 \\
UKIDSS {\em{Y}}   & 0.324 & 1.04 & 1.40 \\
UKIDSS {\em{J}}    & 0.255 & 1.04 & 1.45 \\
UKIDSS {\em{H}}    & 0.176 & 0.910 & 1.45 \\  
UKIDSS {\em{K}}    & 0.373 & 1.29 & 1.79 \\
{\it{WISE}} W1     & 0.120 & 0.548 & 0.910 \\
{\it{WISE}} W2     & 0.0675 & 0.360 & 0.623 \\
{\it{WISE}} W3     & 0.0217 & 0.161 & 0.193 \\
{\it{WISE}} W4     & 0.0867 & 1.02 & 1.36 \\
PACS 70     & 0.0231 & 0.657 & 8.91 \\
PACS 100    & 0.0237 & 0.343 & 5.67 \\
PACS 160    & 0.0193 & 0.259 & 3.13 \\
SPIRE 250   & 0.143 & 2.26 & 18.2 \\
SPIRE 350   & 0.389 & 9.22 & 17.7 \\
SPIRE 500   & 0.373 & 4.54 & 8.08 \\
SCUBA-2 850  & 2.66 & 6.54 & 8.38 \\[1.5ex]
\hline
\end{tabular}
\end{table}

In addition to these continuous diagnostics, we define a set of surface brightness thresholds that are intended to be directly applicable in scientific analyses. For each band, we determine the surface brightness levels $I_{50}$, $I_{75}$, and $I_{90}$ such that, among pixels with surface brightness greater than or equal to $I_p$, a percentage $p$ of the pixels satisfies $R < 0.1$. These thresholds therefore indicate the brightness above which 50\%, 75\%, or 90\% of pixels are expected to be converged to better than the criterion $R<0.1$. The resulting values are reported in Table~{\ref{tab:intensity_percentiles}} and provide a convenient, filter-dependent guideline for identifying regions of the synthetic images where Monte Carlo noise is unlikely to dominate the signal. 

It is important to emphasise that these thresholds should be interpreted as empirical and probabilistic guidelines rather than as strict limits. The precise behaviour of $R$ depends on galaxy morphology, dust distribution, and viewing geometry, and residual uncertainties associated with sparse sampling of the primary stellar sources or with physical modelling assumptions are not captured by the Monte Carlo diagnostics alone. Nevertheless, the thresholds offer a useful first-order indication of the surface brightness regime in which pixel level measurements are expected to be robust against Monte Carlo noise.

Users should also note that the effective signal-to-noise can be increased by spatial binning or downsampling of the images. Because the dominant source of random uncertainty is Monte Carlo sampling noise, binning $n$ independent pixels increases the signal to noise ratio approximately as $\sqrt{n}$, provided that the binned pixels probe comparable surface brightness levels. Spatial averaging therefore provides a straightforward way to extend reliable measurements to lower surface brightness regimes, at the expense of spatial resolution. This trade-off can be tailored to the requirements of a given scientific application. Finally, we note that, for studies on the faint outer regions of galaxies, one can apply biasing techniques in the Monte Carlo postprocessing to boost photon emission from low-density regions \citep{Baes2025b}. This feature is not used in the present version of the TSA, but is available in {\texttt{SKIRT}}. 

\section{Summary, applications, and caveats}
\label{Summary.sec}

\subsection{Summary of TSA DR2}

In this paper we have presented the second data release of the TNG50-SKIRT Atlas (TSA DR2), a comprehensive set of synthetic data cubes, broadband images, and physical property maps for 1154 galaxies drawn from the TNG50 cosmological hydrodynamical simulation and post-processed with the Monte Carlo radiative transfer code {\texttt{SKIRT}}. Compared to the first release, TSA DR2 represents a substantial advance in both scope and robustness.

The most important improvements of TSA DR2 relative to TSA DR1 can be summarised as follows. 
\begin{itemize}
\item The wavelength coverage has been significantly extended by including thermal emission, now spanning the full UV to submm range. This enables direct investigation of dust attenuation and dust emission within a single, self-consistent radiative transfer framework. 
\item The treatment of stellar emission has been updated, incorporating improved template libraries for both old and young stellar populations, resulting in more realistic SEDs and colours across the full wavelength range. 
\item Several technical and methodological issues identified in TSA DR1 have been resolved, most notably the dust allocation artefact discussed in Sect.~{\ref{DustAllocation.sec}}, which affected a subset of galaxies and led to unphysical dust distributions and attenuation signatures. The revised dust allocation scheme implemented in TSA DR2 ensures mass conservation and eliminates these artefacts.
\end{itemize}

In addition to spectral data cubes, broadband images and galaxy property maps, TSA DR2 provides a catalogue of integrated galaxy properties for all galaxies and multiple viewing orientations. These include global physical quantities as well as integrated photometry in commonly used broadband filters. As demonstrated in Sect.~{\ref{IntegratedProperties.sec}}, the integrated luminosities exhibit physically sensible and internally consistent correlations with the instantaneous SFRs of the simulated galaxies, providing both a validation of the radiative transfer calculations and practical guidance for users.

\subsection{Intended use and scientific applications}

The primary purpose of TSA DR2 is to provide a flexible and internally consistent forward-modelling data set that enables controlled investigations of how intrinsic galaxy properties translate into observable quantities across wavelengths and spatial scales. The controlled nature of the simulations and the direct link between intrinsic and observable quantities make TSA DR2 well suited as a training and validation set for machine learning approaches that aim to infer intrinsic galaxy properties from observables, in particular for assessing biases and degeneracies in a controlled setting.

One important application is the study of multi-wavelength SFR tracers. Unlike observational data, TSA DR2 provides direct access to the underlying, instantaneous star formation activity as defined by the simulation, together with matched synthetic observables from the UV to the submm. This makes it possible to investigate how different tracers respond to dust attenuation, dust emission, and viewing geometry. A preview has been given in Sect.~{\ref{IntegratedProperties.sec}}, but we stress that these are intended as diagnostics of internal consistency rather than as empirical calibrations. The current atlas extends such studies to spatially resolved scales, even though this is non-trivial, as discussed by \citet{Kennicutt2012}.

Another important application concerns galaxy dust scaling relations. Relations between dust content, stellar mass, colour, and surface density have been well established on global galaxy scales \citep[e.g.,][]{Cortese2012, Viaene2014, Galliano2021, Abdurrouf2022b}, but there is growing evidence that these relations are driven by local, sub-kpc-scale physics. The work of \citet{Viaene2014} on M31 demonstrated that dust scaling relations observed on galaxy-wide scales closely resemble those found locally within different structural components, while simultaneously revealing a large diversity at the level of individual regions. TSA DR2 provides a unique opportunity to investigate such relations in a statistically significant sample of galaxies, linking global trends to their resolved physical origins within a physically coherent framework.

The atlas also provides a powerful platform for investigating the wavelength dependence of galaxy morphology and structural measurements. The availability of matched dust-free and dust-aware images, together with full UV--submm coverage, allows users to study how dust affects apparent sizes, concentrations, asymmetries, and other non-parametric morphological indicators across wavelength. Previous work has explored this topic using limited wavelength ranges \citep{BokonaTulu2026} or smaller simulation samples \citep[e.g.,][]{Kapoor2021, Camps2022}. With TSA DR2, such analyses can now be extended seamlessly from the UV to the submm, complementing observational studies of multi-wavelength morphology \citep{MunozMateos2009b, Baes2020c} and enabling direct comparison between intrinsic and dust-affected structural properties. 

Beyond these examples, we anticipate that TSA DR2 will find broad use in the community.

\subsection{Scope, limitations, and outlook}

While TSA DR2 offers a rich and versatile data set and represents a significant improvement over DR1, it is important to clearly delineate its scope and limitations.

First, all results presented in the atlas are conditional on the underlying TNG50 simulation and its subgrid physics, including prescriptions for star formation, feedback, chemical enrichment, and the interstellar medium. The synthetic observables should therefore be interpreted as forward-modelled representations of one specific simulated galaxy population, rather than as universally applicable predictions. Applying the same radiative transfer pipeline to a different simulation would likely yield systematically different results.

One specific aspect of the TNG50 simulation that should be taken into consideration is that the TNG galaxy evolution model was calibrated at the resolution of the TNG100 simulation. The higher resolution of TNG50 leads to galaxies that are, on average, somewhat more massive and more actively star-forming than observed systems \citep[e.g.,][]{Pillepich2018a, Pillepich2018b, Donnari2019, Trcka2022, Gebek2024}. These offsets propagate into the synthetic observables generated in TSA DR2.

Another limitation is that the TNG model does not explicitly track dust as a separate component. The dust distribution must therefore be inferred from the gas properties using a sub-grid prescription. While the revised dust allocation scheme in TSA DR2 resolves specific artefacts present in DR1, the overall dust content and distribution remain model-dependent. Moreover, the dust attenuation and emission modelling relies on a fixed dust model ({\texttt{THEMIS}}). While these choices are well motivated and widely used, uncertainties associated with dust composition, emissivity, and PAH properties are not explored here and are not captured by the Monte Carlo uncertainty analysis. Systematic uncertainties related to dust physics potentially dominate over stochastic Monte Carlo noise. Galaxy formation simulations that include a live dust model, such as SIMBA \citep{Dave2019}, NewCluster \citep{Byun2025, Han2026} and COLIBRE \citep{Schaye2026}, indicate that the dust-to-metal ratio is not constant among and within galaxies. 

A final word of caution related to the TNG50 simulation is the resolution. Even though TNG50 has one of the highest resolutions among large-volume galaxy formation simulations, its spatial resolution remains limited. While TSA DR2 provides images at high spatial resolution, the physical interpretation of structures on the smallest spatially resolved scales requires caution. Star-forming regions are modelled using subgrid templates rather than being resolved directly, and the cold ISM structure is not explicitly captured by the simulation. As a result, the atlas is best suited for statistical and comparative analyses on spatially resolved scales, rather than for detailed modelling of individual star-forming regions or fine-grained dust structures. Spatial binning or averaging may be appropriate for quantitative analyses in low-signal regimes.

The limitations of TSA DR2 are not only rooted in the underlying TNG50 simulation, but also in the radiative transfer post-processing methodology. The latter introduces an additional layer of modelling assumptions that are not unique. Different choices for stellar population synthesis models, dust models, or sub-grid treatments of star-forming regions can lead to systematically different synthetic observables, even when applied to the same underlying simulation. In that sense, the {\texttt{SKIRT}}-based pipeline should be regarded as one particular realisation of a broader forward-modelling framework, rather than as a unique mapping between intrinsic and observable properties. This also implies that discrepancies between TSA DR2 predictions and observational data cannot be unambiguously attributed to either the underlying TNG50 simulation or the radiative transfer modelling. The two components are intrinsically entangled in the final synthetic products, and disentangling their respective contributions requires either controlled variations of the post-processing assumptions or comparisons across different simulation and modelling pipelines.

One aspect not yet included is the absence of emission from active galactic nuclei (AGN). Although TNG50 includes black hole particles, their radiative output has not yet been incorporated into the {\texttt{SKIRT}} post-processing. The {\texttt{SKIRT}} code itself supports AGN emission and has been used extensively to model AGN tori and their emission from the optical to the FIR and X-ray regimes \citep[e.g.,][]{Stalevski2012, Stalevski2016, VanderMeulen2023, ReyesAmador2025}. We are currently developing a new AGN template library that self-consistently covers the X-ray to radio wavelength range, including both continuum and line emission, with the aim of incorporating AGN emission in future applications.

A further aspect not yet included is emission from diffuse ionised gas outside star-forming regions. In TSA DR2, nebular emission is included for star-forming regions through the stellar templates, but emission from diffuse gas is neglected. We are developing a new interstellar medium model, inspired by photoionisation calculations with {\texttt{Cloudy}}, that self-consistently calculates the ionisation state and the resulting emission of diffuse gas \citep{Kapoor2026}. At the same time, the {\texttt{TODDLERS}} framework is constantly being updated. The latest version, {\texttt{TODDLERS}} 2.0, supports flexible initial mass functions, stochastic stellar population sampling, a range of cloud density profiles, and configurable dust grain size distributions \citep{Kapoor2026b}.

Finally, while TSA DR2 extends the atlas to the submm regime and full data cubes are generated, the spectra still have relatively modest spectral resolution that is primarily suited to generate broadband images. A natural next step will be to move towards full spectral resolution synthetic data products, similar in spirit to previous efforts \citep[e.g.,][]{Bottrell2022, Nanni2022, Nanni2023, Sarmiento2023}, but with a fully self-consistent treatment of dust attenuation and emission across the entire ultraviolet to millimetre wavelength range. Several building blocks required for this ambition are already in place, including high-resolution stellar and nebular templates, full kinematic support in {\texttt{SKIRT}} \citep{Camps2020, BarrientosAcevedo2023}, and ongoing development of multi-phase ISM sub-grid models.

Despite these limitations, TSA DR2 represents a significant step forward in the availability of physically motivated, multi-wavelength synthetic galaxy data. By combining high-resolution cosmological simulations with state-of-the-art radiative transfer modelling, the atlas provides a valuable bridge between theory and observation.

\begin{acknowledgements}
M.B., N.A., A.G., I.K., A.L., K.M., and W.S. acknowledge financial support by the Flemish Fund for Scientific Research (FWO-Vlaanderen) through the research projects G037822N and G061825N, PhD Fellowship Grant 1193525N (A.L.), and Junior Postdoctoral Fellowship grant 12AJQ26N (K.M.). M.B., P.V., A.G., A.U.K., and A.N. acknowledge funding from the Belgian Science Policy Office (BELSPO) through the PRODEX projects 4000147090, 4000143347, and 4000151690. M.S. acknowledges support by the Ministry of Science, Technological Development and Innovation of the Republic of Serbia (MSTDIRS) through contract No.~451-03-33/2026-03/200002 with the Astronomical Observatory. J.F. acknowledges financial support from the DGAPA-PAPIIT project IN102226, Mexico. S.B.T. and A.T.E. gratefully acknowledge financial support from the NASCERE project, a bilateral cooperation program between Jimma University (Ethiopia) and Ghent University (Belgium).

This study made extensive use of the Python programming language, especially the {\tt{numpy}} \citep{Harris2020}, {\tt{matplotlib}} \citep{Hunter2007}, {\tt{pandas}} \citep{McKinney2010}, and {\tt{Astropy}} \citep{AstropyCollaboration2013} packages. The IllustrisTNG simulations \citep{Marinacci2018, Naiman2018, Nelson2018, Pillepich2018b, Springel2018} were undertaken with compute time awarded by the Gauss Centre for Supercomputing (GCS) under GCS Large-Scale Projects GCS-ILLU and GCS-DWAR on the GCS share of the supercomputer Hazel Hen at the High Performance Computing Center Stuttgart (HLRS), as well as on the machines of the Max Planck Computing and Data Facility (MPCDF) in Garching, Germany. The IllustrisTNG data used in this work are publicly available at \url{https://www.tng-project.org/}, as described by \citep{Nelson2019a}.
\end{acknowledgements}

\bibliography{mybib.bib}

\end{document}